\documentclass[]{aa}
\usepackage{psfig}

\begin{document}
\def\ntilde{\hbox{\rm n}}
\def\vv{\hbox{\bf v}}
\def\gvec{\hbox{\bf g }}
\def\rvec{\hbox{\bf r }}
\def\svec{\hbox{\bf s }}
\def\vvec{\hbox{\bf v }}
\def\solphys{\hbox{Sol. Phys.}}
\def\aaps{\hbox{A\&AS}}
\def\aap{\hbox{A\&A}}
\def\apj{\hbox{ApJ}}
\def\apjl{\hbox{ApJL}}
\def\nat{\hbox{Nature}}

\titlerunning{Multi-wavelength time-resolved coronal loop}
\authorrunning{Reale \& Ciaravella}

\title{Analysis of a multi-wavelength time-resolved observation of a coronal loop}
\author{Fabio Reale\inst{1}, 
\and
Angela Ciaravella\inst{2}}
\institute{Dipartimento di Scienze Fisiche \& Astronomiche, Sezione di
Astronomia, Universit\`a di Palermo, Piazza del Parlamento 1,
I-90134 Palermo, Italy
\email{reale@astropa.unipa.it}
\and
INAF - Osservatorio Astronomico di Palermo ``Giuseppe
S. Vaiana", Piazza del Parlamento 1, I-90134 Palermo, Italy
\email{ciarave@astropa.unipa.it}
}


\abstract{Several items on the diagnostics and interpretation of coronal loop
observations are under debate.} 
{In this work, we analyze a well-defined loop
system detected in a time-resolved observation in several spectral bands to
study how far one can go in characterizing the loop structure and evolution.  }
{The dataset includes simultaneous sequences of images in the 171 \AA, 195
\AA~and 284 \AA~filter bands of TRACE, and in one filter of Yohkoh/SXT, with a
time coverage of about 2.5 hours, and two rasters taken with SoHO/CDS in twelve
relevant lines, forming between $\log T \approx 5.4$ (O V 629 \AA) and $\log T
\approx 6.4$ (Fe XVI 360 \AA). The loop is initially best visible in the TRACE
195 \AA~filter band, with some correspondence with the simultaneous SXT images,
and later in the 171 \AA~filter band, with good correspondence with the CDS
raster images in the lines with formation temperature around $\log T \approx
6.0-6.1$.  We have taken as pixel-by-pixel background the latest TRACE, Yohkoh
and CDS images where the loop has faded out. We examine the loop
morphology evolution, the light curves, the TRACE filter ratio distribution and
evolution, the images and emission measure from the CDS spectral lines.}
{Our analysis detects that, after background subtraction, the emission along the
loop and its evolution are non-uniform, especially in the 171 \AA~filter band,
and that the TRACE 195/171 filter ratio has a moderately non-uniform
distribution along the loop and evolves in time. Both the light curves and the
filter ratio evolution indicate a globally cooling loop. Relatively hot plasma
may be present at the beginning while, during the first CDS raster,
the data indicate a rather moderate thermal structuring of the loop.} {Our data
analysis supports a coherent scenario across the different bands and
instruments, points out difficulties in
diagnostic methods and puts quantitative basis for detailed forward modeling.}

\maketitle

\section{Introduction}

As pointed out by the first X-ray images at high resolution  (e.g. Vaiana et
al. 1973), coronal loops are the building blocks of the X-ray luminous solar
corona. Hot coronal loops are known to be steady and stable over time scales
longer than the characteristic plasma cooling times, and equilibrium scaling
laws ruling some physical conditions (i.e. the maximum temperature, the
pressure and the heating rate) generally hold (Rosner et al. 1978).

The high space and time resolution of the {\it Transition Region And
Coronal Explorer} (TRACE, Handy et al. 1999) telescope has brought new
insight into the structure of the solar corona, and allowed to address
new issues driven by the detection, for instance, of the filamentary
structure of the coronal loops (Schrijver et al. 1999, Reale \& Peres
2000) and of oscillations and propagating waves (e.g. De Moortel et al.
2000, Nakariakov \& Ofman 2001).

In parallel to such achievements, the interpretation of ``conventional" loop
structures as observed with TRACE is under debate.  To summarize, already the
first observations with TRACE detected steady loops which showed to have a
ratio of the fluxes in the 195 \AA~and 171 \AA~filter passbands almost constant
along the loop (Lenz et al. 1999). The ratio of the emission in different
passband filters is typically used as a temperature indicator.  Unfortunately
the temperature diagnostics using TRACE filters are particularly difficult,
because the functions linking the ratios to the temperature are multi-valued.
Nevertheless, if one assumes that most of the emission comes from plasma in the
temperature range of maximum filter sensitivity, the function can be inverted
and what is typically obtained for TRACE loops is that they are almost
isothermal, much more than predicted by standard static loop models, and also
overdense with respect to static loops at $\sim 1$ MK.  Was it a new class of
loops?  Soon after, it was shown that an alternative interpretation is
possible: bundles of thin strands, each behaving as a single ``standard" loop,
convolved with the TRACE temperature response could appear as a single almost
isothermal loop (Reale \& Peres 2000). Next to this, another possibility has
been invoked that long loops result to be mostly isothermal if heated at their
footpoints (Aschwanden et al.  2001). This model suffers from the problem that
footpoint-heated loops have been proven to be thermally unstable (e.g. Serio et
al. 1981)  and therefore cannot be long-lived, as instead observed. A further
alternative is to explain observations with steady non-static loops, i.e. with
significant flows inside (Winebarger et al. 2001, 2002). Also this hypothesis
does not seem to answer the question (Patsourakos \& Klimchuk 2004). One of the
reasons why the situation is so unclear is inherent in the data: although
sometimes bright and well defined, the loops under analysis are always
surrounded by other bright structures, which often intersect them along the
line of sight.  Moreover, a uniform diffuse background emission also affects the
temperature diagnostics, by adding systematic offsets which alter the filter
ratio values. This problem emerged dramatically when the analysis of the same
loop structure observed with the {\it Soft X-ray Telescope} (SXT, Tsuneta et
al. 1991) on board Yohkoh (Ogawara et al. 1991) led to three different results
depending mostly on the different ways to treat the background (Priest et al.
2000, Aschwanden 2001, Reale 2002). One obvious way to check for the validity
of the data analysis and of the loop diagnostics is to compare data from
imaging instruments to simultaneous and cospatial data obtained from
spectroheliographic instruments like the {\it Coronal Diagnostic Spectrometer}
(CDS, Harrison et al. 1995) on board SoHO (Domingo et al. 1995).  A loop
observed on the solar limb with SoHO/CDS was analyzed by Schmelz et al. (2001),
who found that whereas single line ratios tend to yield flat temperature
distributions along the loop, a careful reconstruction of the emission measure
distribution vs temperature (DEM) at selected points along the loop shows that
this may not be a realistic result. A whole line of works started from this
study reconsidering and questioning the basic validity of the temperature
diagnostics with TRACE and emphasizing once again the importance of the
background subtraction, but also the need to obtain accurate spectral data
(Schmelz 2002, Martens et al. 2002, Aschwanden 2002, Schmelz et al. 2003).
Similar results but different conclusions are reached by Landi \& Landini
(2004), and Landi \& Feldman (2004) who analyze a loop observed with SoHO and,
finding it nearly isothermal, consider this evidence as real and invoke a
non-constant cross-section to explain it. From their analysis of SoHO/CDS data
compared to other similar analyses made by other authors, Schmelz et al. (2005)
propose that  there may be two different classes of loops, multi-thermal and
isothermal, while Aschwanden \& Nightingale (2005) analyze the thinnest loop
structures detected with TRACE and find that a few are isothermal along the
line of sight and may therefore be elementary loop components.

Another puzzling issue, certainly linked to the loop isothermal
appearance, is the loop overdensity. In order to explain both these
pieces of evidence, several authors claim that the loop cannot be at
equilibrium and it must be filamented and cooling from a hotter state,
probably continuously subject to heating episodes (nanoflares, Warren
et al. 2002, Warren et al. 2003, Cargill \& Klimchuk 2004).  The
presence of nanoflares might explain the presence of coronal loops,
stable although heated at the footpoints and with a peaked distribution
of emission measure, as observed in active stars (Testa et al. 2004).
An anticoincidence between hot and cooler loops has been found from the
comparison of simultaneous Yohkoh and TRACE data (Nagata et al. 2003, Schmieder
et al. 2004), who, however, find that the DEM of loops have a moderate
but finite width.

Time-dependent modeling of one coronal loop observed with TRACE pointed
out that the detailed description of the evolution of this loop
requires a heating located at intermediate position between the apex
and the footpoints, probably initially high and then slowly decaying
(Reale et al. 2000).

The current debate in the interpretation of coronal loop observations points
out the presence of intrinsic limitations in the  information that one can
derive from present-day data.  Here, we take the analysis of a multi-wavelength
observation of a time-evolving coronal loop as a guide to study how deep one
can go in the  diagnostics and characterization of the loop, and puts  the
basis for further analysis through detailed forward modeling which we leave for
a future work. To this purpose we have searched for the observation of a loop
in particularly good conditions for analysis: a simple and well-defined system,
as isolated as possible, imaged in more than one TRACE filter band, in several
SoHO/CDS spectral lines and with Yohkoh/SXT, and for a time period of more than
one hour. Its eventual disappearance allows us to use the last images as
point-to-point background to be subtracted. We try to use the coherence of the
structure, its evolution and the spectral data to extract the maximum possible
information from the data.

The selection of the loop observation, its description and the methods
of the data analysis are illustrated in Section~\ref{sec:data}, the
results of the analysis are shown in Section~\ref{sec:res} and they
are discussed in Section~\ref{sec:disc}; we draw our conclusion in
Section~\ref{sec:concl}.

\section{The observation and data analysis}
\label{sec:data}

\subsection{The observation campaign}

\begin{figure}
\centerline{\psfig{figure=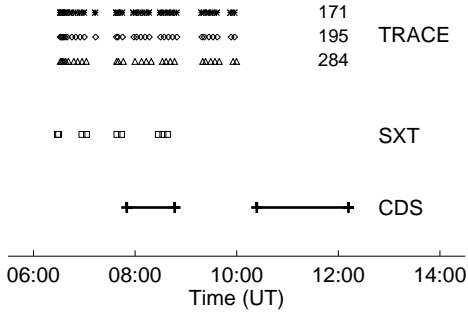,width=8cm}}
\caption[]{Time coverage of the relevant observations of
the loop selected for the present analysis. The times of
the TRACE and Yohkoh frames are marked with data points ({\it stars, diamonds
triangles, and squares} for the TRACE 171 \AA, 195 \AA~and 284 \AA~filters and for
SXT, respectively), the time lapses of the two relevant CDS rasters are
marked with the segments.
\label{fig:times}}
\end{figure}

From the list of SoHO campaigns available at the Web site
\underline{http://sohodata.nascom.nasa.gov/cgi-bin/gui}, we have selected the
campaign, with ID number 5170, named {\it Active Region Study}. The Observation
date is 13 May 1998 and the coordinator is Robert Walsh. The campaign includes
data from TRACE, SoHO/CDS and Yohkoh/SXT. TRACE data consist of a 3.5 hours
time-sequence of 1024$\times$1024 full resolution images in all three filters
(171 \AA, 195 \AA, 284 \AA). Yohkoh/SXT data consist of a time-sequence of
half-resolution (5" pixel size) full-disk images mostly in a single filter,
with a good time overlap with the time period of the TRACE data.  SoHO/CDS data
include several rasters but only two of them with good spectral resolution and
a relevant field of view (see Section~\ref{sec:cdsdata}). The first raster
overlaps the TRACE observation. Fig.~\ref{fig:times} shows the time coverage of
each piece of observation relevant for the present analysis. The TRACE frames
span from 6:30 UT to 10 UT with two significant gaps of about 20 min around
7:15 UT and 8:50 UT, which divide the TRACE data into three main parts.  Most
of the loop evolution is included in the first two TRACE segments. These are
well covered by the SXT data (until 8:50 UT). The first CDS raster occurs
during the second TRACE segment. The second (longer) raster is taken after the
end of the TRACE data. For completeness, in order to report on the loop
evolution at times before the campaign, we have partially analyzed TRACE data
in the 171 \AA~ and 195 \AA~filter bands (about 50 additional images in each
filter) taken between 0:18 UT and 6:20 UT and one Yohkoh/SXT image taken at
5:26 UT.

\subsection{The loop selection}

The loop has been selected on the TRACE images. It appears as an
entire loop in several 171 \AA~filter images, and it is clearly
visible also in the 195 \AA~band. A mostly visible loop is
well suitable to provide complete constraints on diagnostics and
models. Many loops are faint around their apex, simply because of the
stratification due to gravity. The loop is bright, i.e. observed with
good count statistics and with a high contrast over the background.
We have searched for a loop as far as possible free from other
structures intersecting the line of sight, which might be difficult
to disentangle from the analyzed loop. The right half of the loop
appears to match well this requirement. The loop is  $\approx 10^{10}$
cm long and lies on the solar disk ($X_{sol}= +230"$, $Y_{sol}=+470"$)
(see also Fig.~\ref{fig:fov}). We prefer a loop located on the disk,
because, on the limb, more coronal structures are met along the line of
sight (as suggested by a simple inspection of TRACE images) and their
contribution is more difficult to subtract.

The multi-wavelength observation allows for a cross-check of the
results and for absolute calibration. Data in multiple TRACE EUV filter
passbands allow for imaging and filter ratio diagnostics, SoHO/CDS data
for spectroscopy, and Yohkoh/SXT for imaging in X-ray hotter passbands.
The good visibility both in the 171 \AA~and in the 195 \AA~filter band
ensures a good S/N ratio for filter ratio (and therefore temperature)
diagnostics.  It is also visible in the 284 \AA~filter band.  Yohkoh/SXT
data are of good quality, although mostly in a single filter band;
a loop light curve can be derived from them.  The SoHO/CDS observation
includes data in several spectral lines, with a good nominal coverage
in the $4.5 < log(T) < 6.5$ temperature sensitivity range.

It is important for our analysis that the data give us information about
the loop evolution. A long-lived and steady loop can be better studied,
because close to physical equilibrium conditions.  Although the selected
loop eventually fades away, its observed intensity evolves slowly,
and is nearly steady in a sequence of several TRACE images. We can then
reasonably assume that the average properties of the loop plasma evolve
slowly as well.

\subsection{The TRACE data analysis}

The campaign dataset includes four relevant TRACE data cubes.  An overall
number of 32 $1024 \times 1024$ images for each filter are available, with
exposure times between 1 and 39 s, between 2 and 46 s, and between 6 and
131 s, for the 171 \AA, the 195 \AA~and the 284 \AA~filter, respectively. In
the 171 filter, the longest exposure time is the most frequent one (22
frames). In the 195 \AA~filter, the exposure time is 28 s for 17 frames,
46 s for 6 frames.  In the 284 \AA~filter, 17 frames are taken with the
longest exposure time.

The data have been processed using the standard IDL procedure {\it
trace\_prep}.  A smaller region ($512 \times 512$ pixels) of the whole
field of view -- the one enclosing the loop -- has been extracted for
analysis.  For each filter band, the images have been coaligned with
cross-correlation between the $512 \times 512$ images. We have removed
the clearly corrupted frame 11 from all three datasets.  In the 284 \AA~
filter frames 10 and 24 are clearly damaged by cosmic rays and
removed.

\subsubsection{Background subtraction}

In order to apply standard diagnostic methods to derive physical quantities,
such as the temperature, or even to apply more detailed loop models, we have to
extract the emission from the loop and exclude any  other contribution along
the line of sight. Background subtraction may not be trivial in a region so
rich in bright structures (e.g. Reale2002, Schmelz et al. 2003). However, we
have taken  advantage of a particularly fortunate situation: the loop
disappears at the end of the image sequence.  The last images (around 10 UT)
can then be used to derive a reliable  background, in the assumption that the
structures surrounding and crossing the selected loop along the line of sight
do not change much during the observation sequence.  The procedure has then
been to simply subtract the last image from all the preceding images. This has
been done for each filter image sequence.

This approach to subtract background has several advantages: it is in
principle very accurate, if the loop under analysis is the structure
which mostly varies in the observation; it is direct, with no use of
interpolation; it is applied pixel by pixel, allowing us to derive
``background-subtracted image" and therefore to have a visual feedback,
and to analyze all loop pixels, instead of sampling at selected
positions. As drawbacks, this method cannot take time-variations
of the background emission into account and we cannot exclude that
crossing structures vary during the observation. We have estimated
the average time fluctuations during the observation by computing
pixel-by-pixel the standard deviation of the count rate. We obtain an
average standard deviation of 13\% both in the 171 \AA~and in the 195 \AA~
filter band (not included in error bars), that we retain a reasonably 
acceptable value that validates
our analysis. Although there is evidence of systematic effect due to
co-evolving nearby structures, the overall validity of this approach is
proven by the results obtained (Sec.~\ref{sec:res}).

\subsection{The Yohkoh/SXT data analysis}

The campaign data consist of two full-frame datasets, with nine relevant
frames (full disk $512 \times 512$ pixels, each with a side of 4.9 arcsec). The
exposure time of all frames is 2668 ms.  Only the first frame is taken
in Al.1 filter band. All others are in the Al/Mg/Mn filter band. The
frames are processed according to the standard Yohkoh procedure {\it
sxt\_prep}. Frames are co-aligned to each other and cross-aligned with
the TRACE images (using the TRACE $512 \times 512$ 284~\AA~ image at 6:57
UT) by means of the cross-correlation method.  As for the background
subtraction, the last frame has been subtracted from the previous ones
in the Al/Mg/Mn filter band, analogously to what we have done with TRACE
data. We will show results only for this passband.

\subsection{The SoHO/CDS data analysis}
\label{sec:cdsdata}

CDS was observing from 07:13:15 to 12:11:58 UT with 15 rasters in the
region of the loop structure. Most of the rasters could not be included
in the present analysis. The field of view of the first raster is not
so large as to include the whole loop. The spectral ranges
of the rasters from the third to the 13$^{th}$ are too narrow to scan the
whole line profiles. Thus, two rasters are useful to our analysis,
those taken between 07:50:09 and 08:46:02 UT (s11107r00) and between
10:23:42 and 12:11:58 UT (s11109r00, Fig.~\ref{fig:times}). The latter
has been used as background, since the loop had already faded out at
that time. The filename, time, coordinates, spatial width and bin size
of both rasters are listed in Table~\ref{tab:cds}.

\begin{table*}
\caption{CDS File Data}
\label{tab:cds}
\begin{center}
\renewcommand{\arraystretch}{0.8}
\begin{tabular}{l c c c c c c c c}
\hline
\hline\\
Filename & $t_i$ & $t_f$ & X$_0$ & Y$_0$ & X-width & Y-width & 
X-bins & Y-steps \\
         & (UT) & (UT) & (") & (") & (") & (") & & \\
\hline
\\
s11107r00 &  07:50:09 & 08:46:02 & 208.0 & 489.3 & 243.8 & 243.8 & 60 & 72\\
s11109r00 & 10:23:42 & 12:11:58 & 261.6 & 489.2 & 243.8 & 243.8 & 120 & 143\\
\\
\hline
\end{tabular}
\end{center}
\end{table*}

The standard CDS software has been used to remove the spikes from the
data and to calibrate them.  Table ~\ref{tab:cdslines} lists the lines
shared by both relevant rasters and used for the present analysis.

\begin{table}
\caption{Observed Spectral Lines}
\label{tab:cdslines}
\begin{center}
\renewcommand{\arraystretch}{0.8}
\begin{tabular}{l c c c}
\hline
\hline\\
Line & Wavelength & $\rm log T$  \\
     &  [\AA]    &  [K] \\
\hline
\\
O V     & 629 & 5.4 \\
Ca X    & 558 & 5.9 \\
Mg IX   & 368 & 6.0 \\
Mg X    & 625 & 6.1 \\
Si X    & 347 & 6.1 \\
Si X    & 356 & 6.1 \\
Fe XII  & 364 & 6.1 \\
Fe XIII & 348 & 6.2 \\
Si XII  & 520 & 6.3 \\
Fe XIV  & 334 & 6.3 \\
Fe XIV  & 353 & 6.3 \\
Fe XVI  & 360 & 6.4 \\
\\
\hline
\end{tabular}
\end{center}
\end{table}

The two rasters have different field of views, number of bins along X
and number of steps along Y (see Table~\ref{tab:cds}).  TRACE and CDS
images have been co-aligned by cross-correlating
the $512\times512$ TRACE 171~\AA~ image at 8:33 UT and the Mg IX 368~\AA~ image 
in the first CDS raster. We have co-aligned the images of the first raster
to those of the second raster line per line.
Following the same procedure as that
used for the TRACE and Yohkoh data, the images of the second raster
have been subtracted from the respective images of the first raster,
for the lines shared by the rasters.

\section{Results}
\label{sec:res}

\subsection{The loop region}
\label{sec:region}

\begin{figure*}
\centerline{\psfig{figure=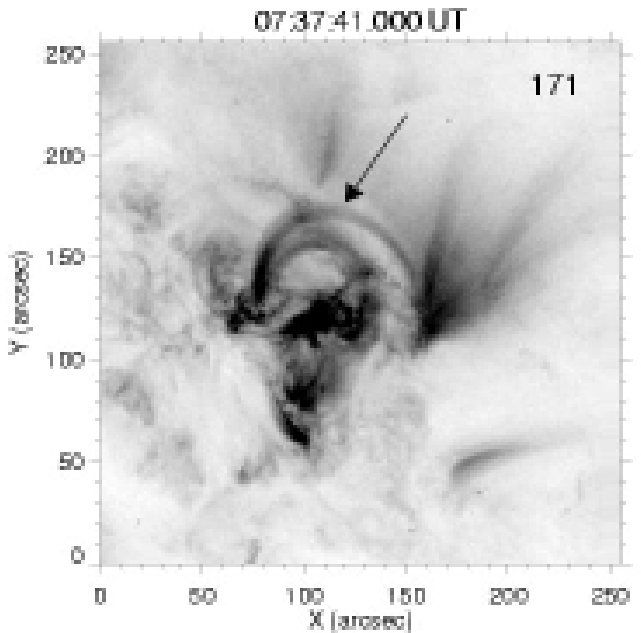,width=12cm}
\psfig{figure=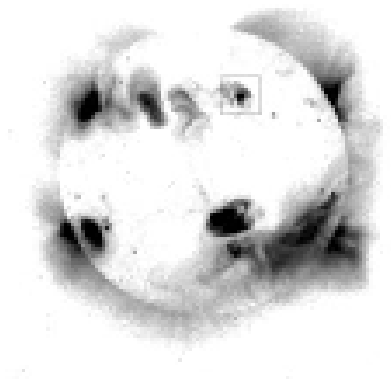,width=6cm}}
\centerline{\psfig{figure=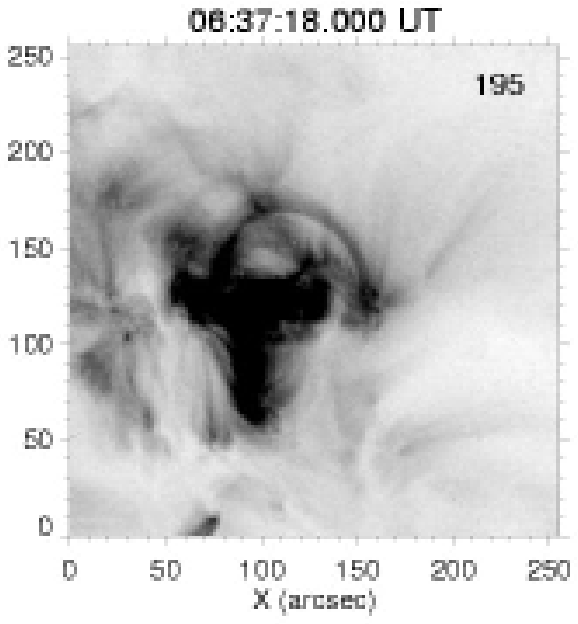,width=6cm}
\psfig{figure=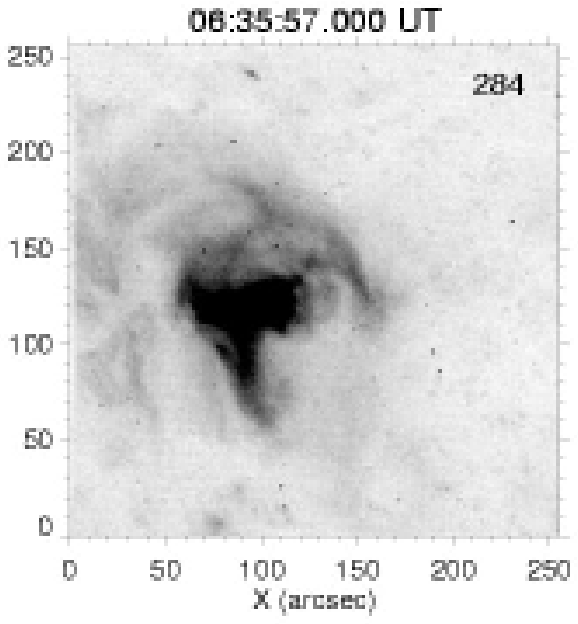,width=6cm}
\psfig{figure=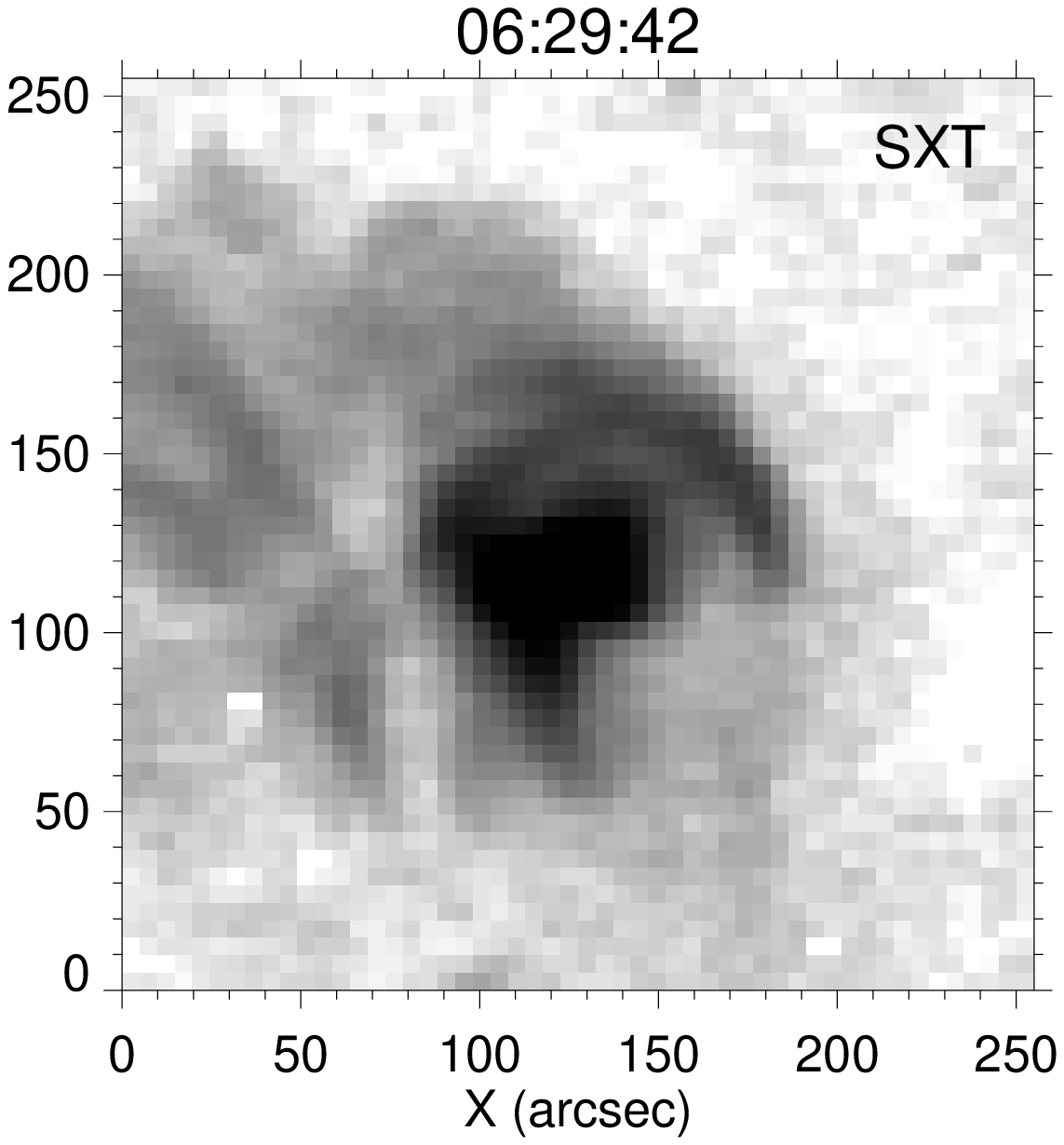,width=6cm}}
\caption[]{The loop region ($256 \times 256$ arcsec) in the passbands
of the 171 \AA, 195 \AA~and 284 \AA~TRACE filters and in the Al/Mg/Mn
sandwich filter of the Yohkoh/SXT. The grey scale is inverted and
linear for all the TRACE images ($\leq 8$ DN/s/pix, $\leq 6 $ DN/s/pix,
$\leq 2 $ DN/s/pix), and logarithmic for the Yohkoh image (between 10
and 150 DN/s/pix). In the 171 \AA~image the loop is indicated with an
arrow. The loop region is located in the inset of the Yohkoh/SXT 
full disk image ({\it upper right}).
\label{fig:fov}}
\end{figure*}

Fig.~\ref{fig:fov} shows the loop region in the TRACE 171 \AA, 195 \AA, 284
\AA~filter passbands and in the Al/Mg/Mn filter passband of Yohkoh/SXT,
and its location on the solar disk.  The loop is very well visible in
the TRACE 171 filter passband and is embedded in a region containing
other bright structures.  Some of them are well-defined, others not.
There is a well-defined smaller concentric loop, which may itself
deserve investigation. In the center of the field of view, there is a
bright moss region, which has weaker filaments southwards. From the
right footpoint of the selected loop, bright plumes mark the base of
larger scale structures whose higher parts are invisible, probably
because higher than the local pressure scale height. Another relatively
bright structure appears to intersect the loop leftward of the apex.
This structure may be really interacting with the loop, as illustrated
below.

In the 195 \AA~filter band, the loop region appears significantly
different from the images in the 171 \AA~filter band, and, in particular,
the central moss region appears far brighter and extended, and the
downward tail more prominent. However, even with no background
subtraction, {\it the loop is clearly visible in the 195 \AA~filter
band}.

In the 284 \AA~good images, the loop region appears more similar to that
in the 195 \AA~filter band than in the 171 \AA~filter band, with a very
prominent central moss region, but generally more diffuse, as
typically occurs in the 284 \AA~filter band. It is worthwhile to remark that
the sensitivity in this filter is more than one order of magnitude
lower than in the 171 \AA~filter band (Handy et al. 1999) and would
require an equivalent increase of exposure time to obtain equivalently
contrasted images. Instead, the exposure time is increased by a factor
3 to 5; the lower S/N ratio may explain, at least in part, the lower
contrast and definition of the structures in the region.
In the initial images, a loop structure similar to those visible in the
other two filter bands is visible, although more diffuse and weaker.
The central moss region is even more dominant in the 284 \AA~images.

In the Yohkoh/SXT Al/Mg/Mn filter passband, the loop region appears
similar to the region in the TRACE 284 \AA~filter band. In particular, we
note the bright feature at the center of the field of view, and the
loop arcade above it.

\subsection{The loop evolution}
\label{sec:evol}

\begin{figure*}
\centerline{\psfig{figure=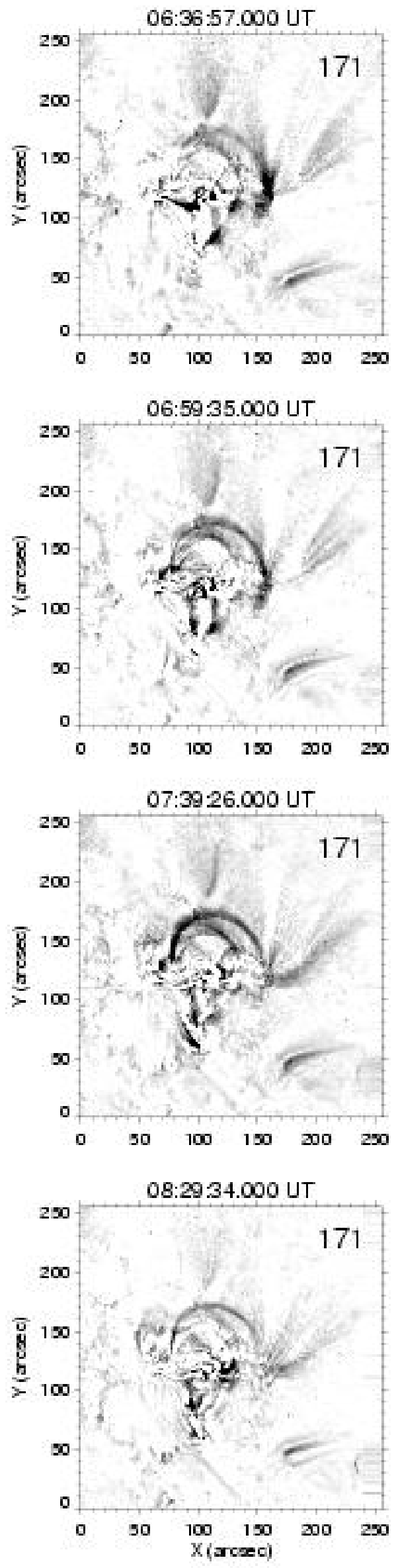,width=4cm}
\psfig{figure=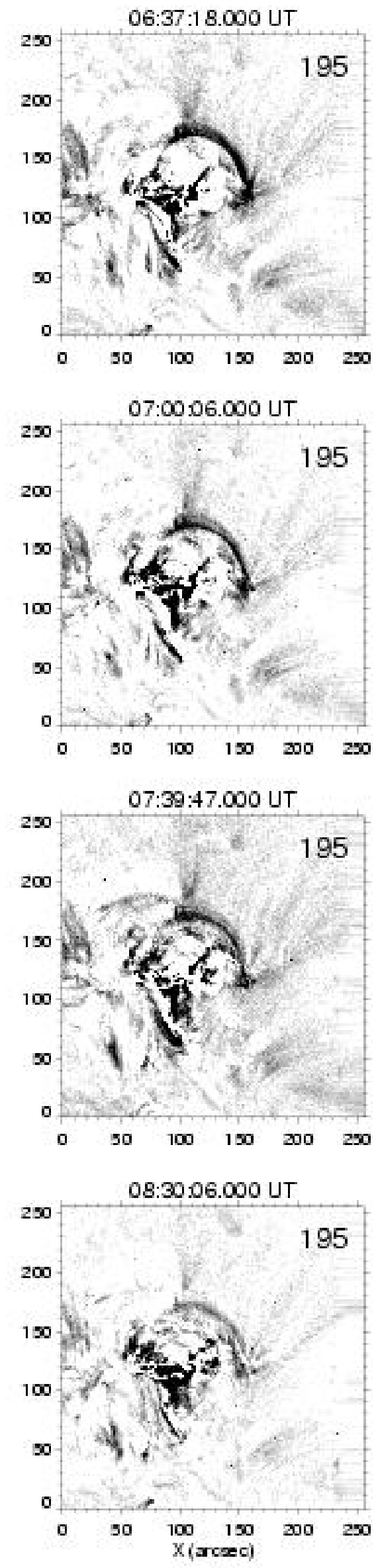,width=4cm}
\psfig{figure=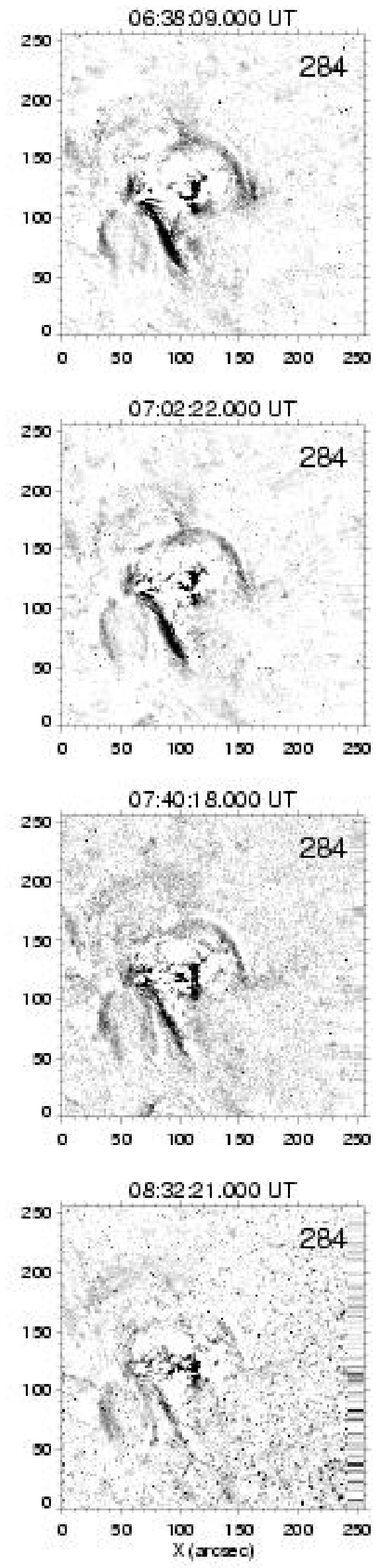,width=4cm}
\psfig{figure=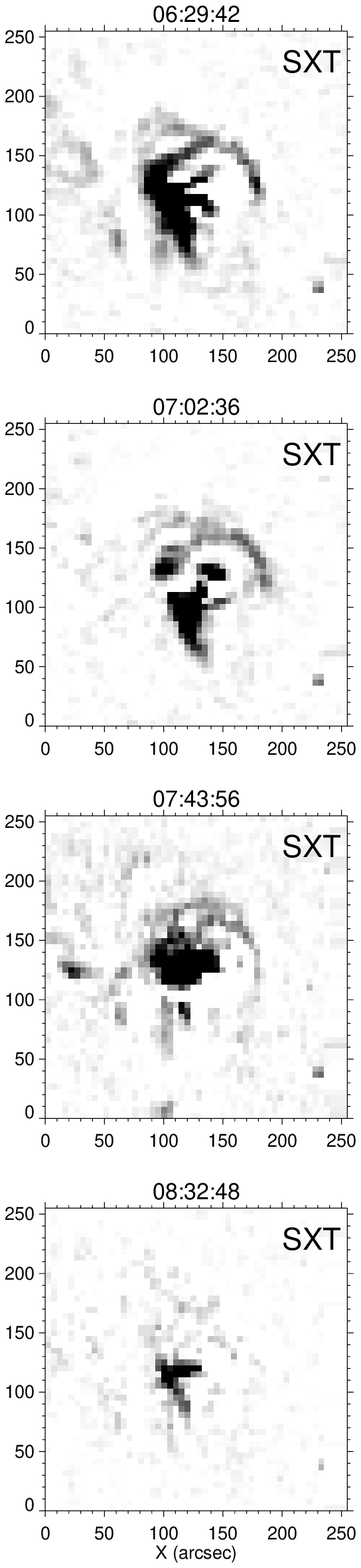,width=4cm}}
\caption[]{Background-subtracted image sequence of the loop region
($256 \times 256$ arcsec) in the passbands of the 171 \AA, 195 \AA~and 284
\AA~TRACE filters and in the Al/Mg/Mn sandwich filter of the Yohkoh/SXT,
from  6:31 UT to 8:31 UT.  The grey scale is inverted and linear for
all the images ($\leq 8$ DN/s/pix, $\leq 6 $ DN/s/pix, $\leq 2 $
DN/s/pix, $\leq 150 $ DN/s/pix).  Negative values have been put to
zero.
\label{fig:fovbk}}
\end{figure*}

The loop evolution during the campaign is summarized in the sequence of
four background-subtracted images shown in Fig.~\ref{fig:fovbk}, in the
TRACE 171 \AA, 195 \AA, 284 \AA~filter passbands and in the Yohkoh/SXT
Al/Mg/Mn filter passband, at corresponding times.  In these images,
all negative values have been put to zero to emphasize the brightness
excesses over the last image. The images show that also other structures
are variable in the loop region, e.g. the central moss region, the inner
loop and other outer loop structures.
The loop is visible with better contrast after background subtraction,
in all TRACE and SXT passbands, in the initial phases of the observation,
and in several CDS lines (see Fig.~\ref{fig:cds1}).  Since it is well
inside the solar disk and its shape well approximates a semi-circle,
it is probably inclined on the surface.

In the TRACE 171 \AA~filter passband, the selected loop is mostly
visible already since the beginning, but it becomes completely visible
half an hour later, at about 7:00 UT, and bright at maximum at 7:30 UT.
It begins to fade significantly about one hour later and is no longer
visible at about 9:15 UT.  In the 171 \AA~images, the right leg of the
loop is initially the brighter.  Later on, the brightness becomes more
uniform along the loop and eventually the left leg becomes the brighter,
before the whole structure fades away.

In the TRACE 195 \AA~filter passband, the evolution is slightly
different, in that the loop appears to be monotonically and uniformly 
fading out toward the end of the observation.  
The loop appears to be truncated quite abruptly on the left side for about 
1/4 of its length, and to be bright again close to the left footpoint. We have
checked that in the dark region the background is particularly intense due to
a bright intersecting structure still present in the last image. The presence 
of this bright structure may affect the background subtraction there.
The right side is initially quite uniformly bright, with a ``halo" on
the outer shell. We have checked that the 171 \AA~loop overlaps the
brightest arch of the 195 \AA~background-subtracted images 
(see also Sec.~\ref{sec:loop} and \ref{sec:filt}).

In the TRACE 284 \AA~filter passband, the loop structure is quite faint
and overlaps only the inner part of the loop of the other two passbands.  
It fades out rapidly. The right leg is quite uniformly bright, the left leg 
has a gap close to the footpoints.

In the Yohkoh/SXT Al/Mg/Mn filter passband, a loop structure appears in
the initial background-subtracted images at a similar location and with
a similar shape as the loop visible in the TRACE passbands, in spite of
the different passband. The right leg of the Yohkoh/SXT loop overlaps
well with the right leg of the TRACE loop. As time progresses, the loop
becomes fainter and fainter, and it is  barely visible at 7:44 UT and
no longer at 8:30 UT. The bright region southward of the loop also
appears to be variable and fading to the end of the observation.

From an inspection of the TRACE images of the loop region before the
beginning of the campaign, we have checked that the loop is present and
bright since about 2 UT (see also Sec.~\ref{sec:lc}).

\subsection{Loop data analysis}
\label{sec:loop}

\begin{figure}
\centerline{\psfig{figure=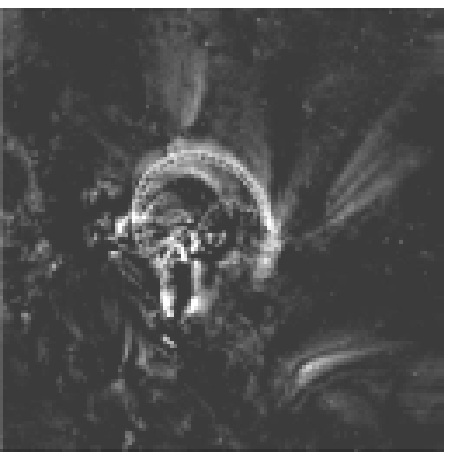,width=5.5cm}}
\centerline{\psfig{figure=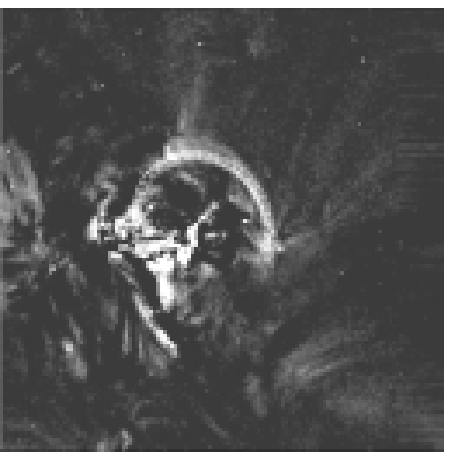,width=5.5cm}}
\centerline{\psfig{figure=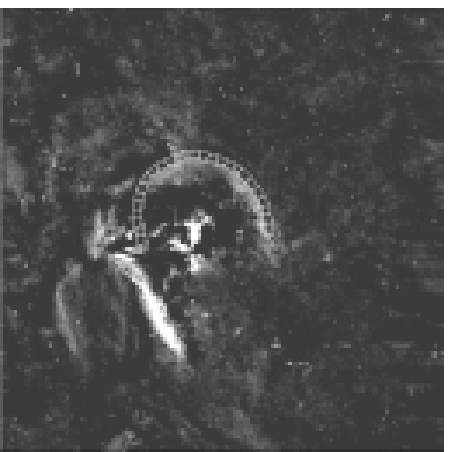,width=5.5cm}}
\centerline{\psfig{figure=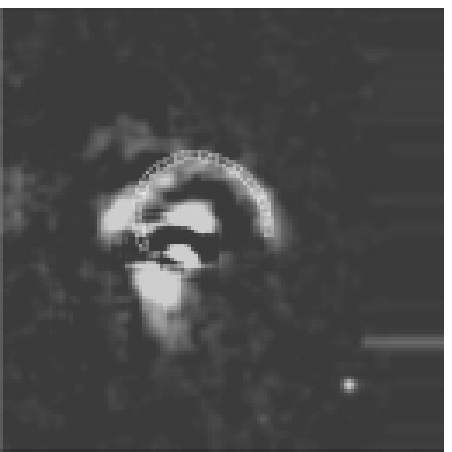,width=5.5cm}}
\caption[]{Counts for loop analysis are extracted from the ({\it white}) 
strip on the images (second row in Fig.~\ref{fig:fovbk} shown with an inverse
gray scale), 10 pixels wide. The count profiles are built after
dividing the strip into sectors (10 pixels long).
\label{fig:loop_sec}}
\end{figure}

Our next step is to analyze the emission inside the loop and the related
diagnostics.  We define a strip enclosing the loop in the TRACE 171
\AA~filter passband, down to the visible footpoints and even beyond
them, and divide it into sectors, as shown in Fig.~\ref{fig:loop_sec}.
We have analyzed strips of different widths; here we show results for
a width of 10 pixels, which encloses the bulk of the loop and represents
a good compromise:  a thinner strip encloses too few pixels for good
enough statistics and may be severely affected by slight alignment errors;
a wider strip may not warrant enough coherence across the loop, i.e. it
may include loop strands in too different conditions to define average
properties. With the choice of a length of 10 pixels, we end up with
a number of 27, almost square, sectors.  Starting from the left extreme
the first three sectors are beyond the visible loop, and the fourth one
partially includes the loop extreme.  The fifth sector can be reasonably
considered as the first visible piece of the loop. On the other side,
we may include all but the last sector.

The DN counts are extracted from each sector, at corresponding locations
in the different passbands of TRACE and Yohkoh/SXT, after cross-aligning
the images, independently of the degree of overlap of the loop structure
in the different passbands (i.e., the bins have the same solar
coordinates in the different passbands to within the coalignment
uncertainties). The SXT images have been rebinned to the same pixel
size of TRACE, so to use the same TRACE strip and sectors for the
emission extraction.

\subsubsection{Profiles along the loop}

\begin{figure*}
\centerline{\psfig{figure=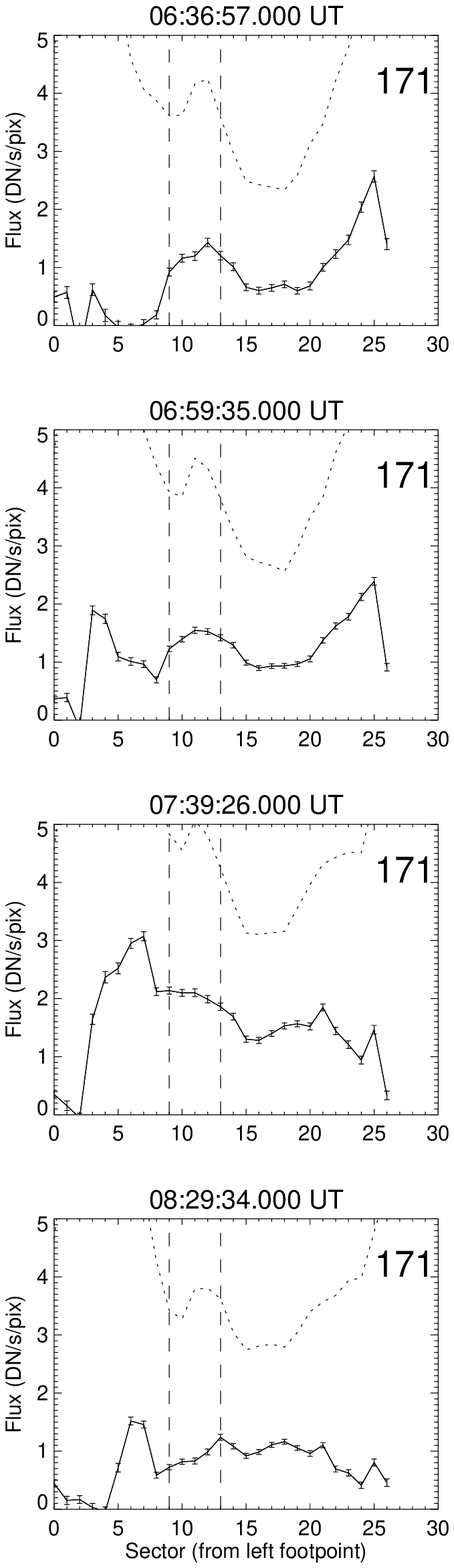,width=4cm}
\psfig{figure=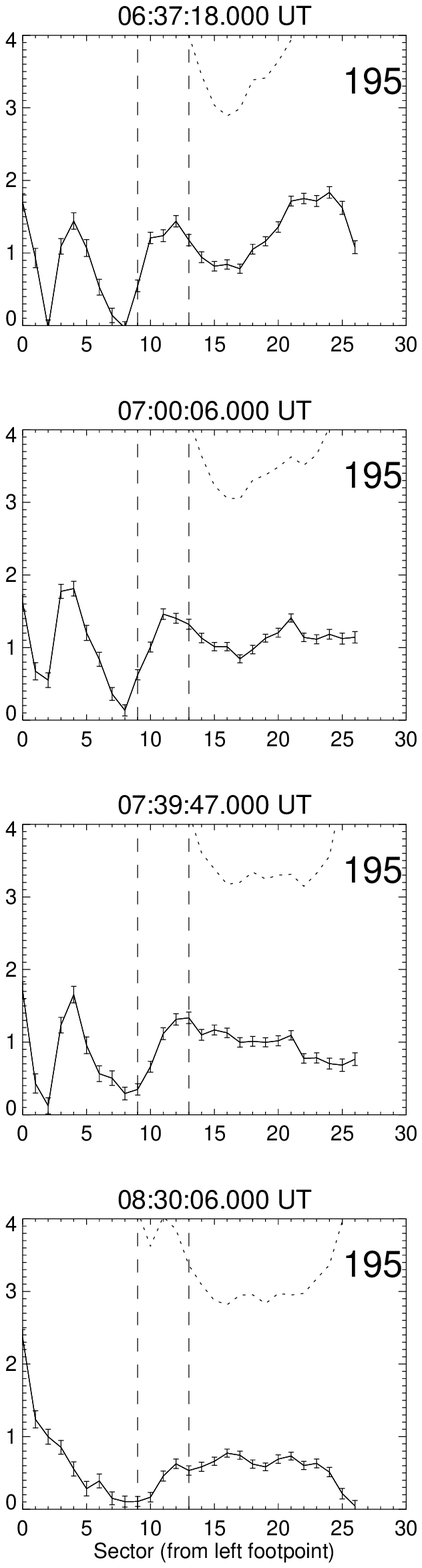,width=4cm}
\psfig{figure=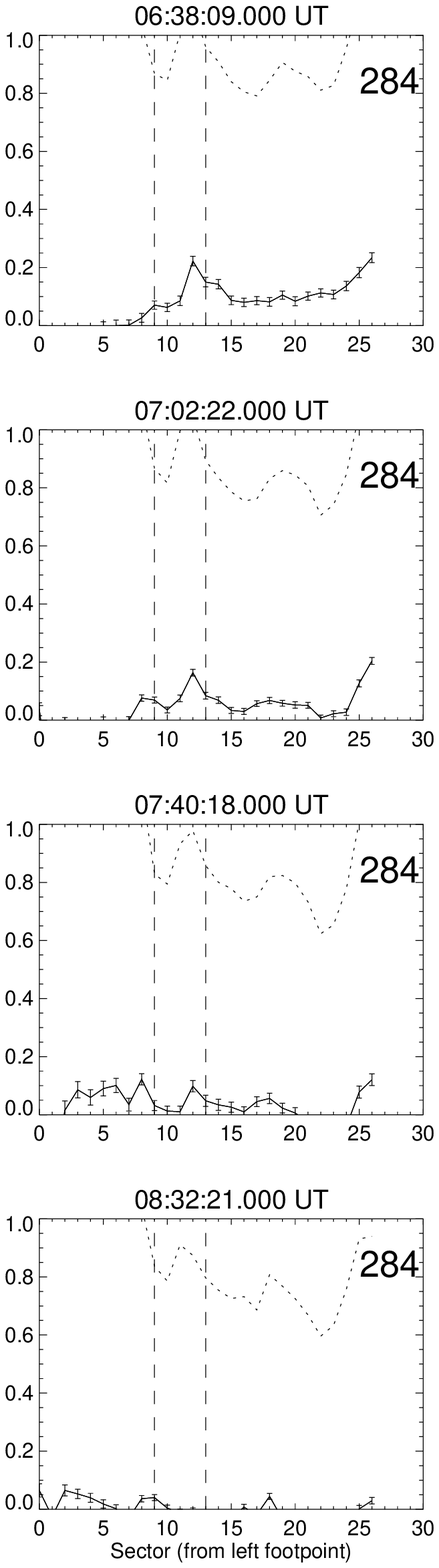,width=4cm}
\psfig{figure=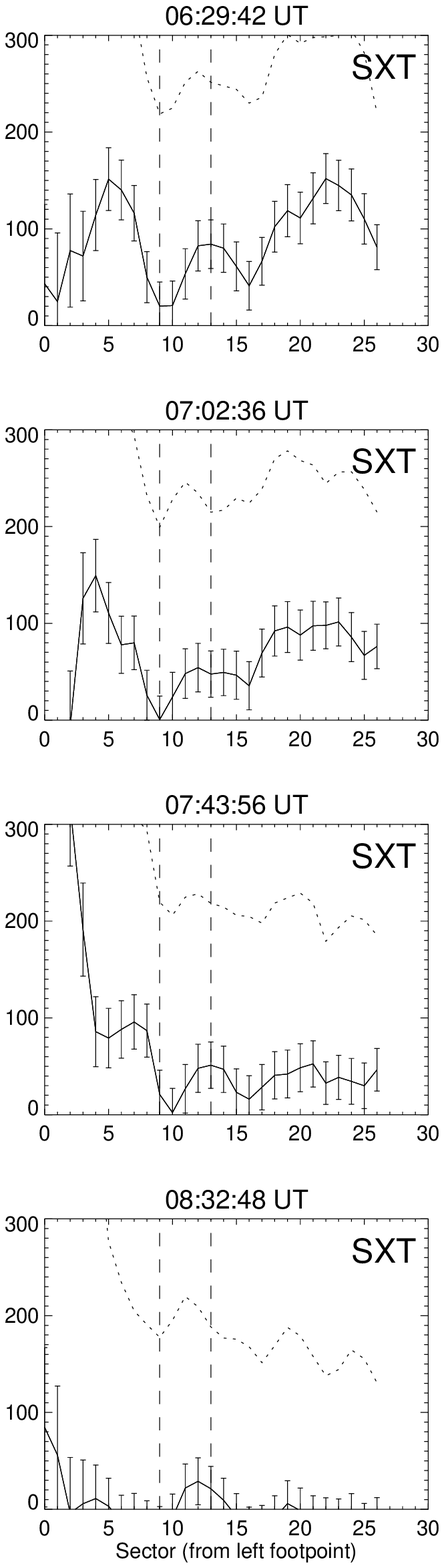,width=4cm}}
\caption[]{Background-subtracted emission profiles ({\it solid, data points})
along the loop in
the 171 \AA, 195 \AA~and 284 \AA~TRACE filter passbands and in the Yohkoh/SXT
Al/Mg/Mn filter. The unsubtracted profiles are also shown (dotted lines). The
dashed vertical lines bound the region in which a bright irregular
structure intersects the loop along the line of sight. 
\label{fig:profbk}}
\end{figure*}

Fig.~\ref{fig:profbk} shows the background-subtracted emission
profiles along the loop in the three TRACE filter passbands and
in the Yohkoh/SXT Al/Mg/Mn filter passband at the same times as in
Fig.~\ref{fig:fovbk}. The error bars are computed as the sum in
quadrature of the error on the total photon counts and of the error
on the background photon counts. The dashed vertical lines mark the
approximate boundaries of the bright structure which intersects the loop
(see Sect.~\ref{sec:region}). This structure produces a clear bump in
most of the emission profiles along the loop (and in all passbands);
since it appears to evolve together with the loop, it probably interacts
with the loop.  However, it should not be part of the loop and we will
exclude it from the loop analysis.  We see in Fig.~\ref{fig:profbk} that
the global emission level is much lower than that of the unsubtracted
profiles in all passbands -- of the order of a half or even less --
indicating that the background emission is indeed a significant fraction
of the total emission from the region.

At the beginning, the loop emission comes mostly from its right leg
in the 171 \AA~filter passband, and is more uniform in the 195 \AA~and 284
\AA~filter passbands. In the Yohkoh/SXT initial profile the emission is
significant in the loop legs and low around the apex.
The emission profiles confirm that, in the 171 \AA~filter passband, the
emission evolution is asymmetric: the left leg of the loop
progressively brightens and the right leg correspondingly fades.  At
7:00 UT the emission profile is almost flat along the loop.  At 7:40 UT
the left extreme is bright and the other is dark. Then also the left
leg becomes fainter, and after 8:30 UT the whole loop is virtually dark,
although there is low and decreasing residual emission. The bump
becomes less and less prominent with time and in the second and third
part of the observation it becomes just a slight change of slope in
the emission profiles.

The background-subtracted 195 \AA~filter profiles show a dip
on the left side (between sectors 5 and 10). This feature is also present
in the background-subtracted images (Fig.~\ref{fig:fovbk}) and probably
due to a co-evolving intersecting structure (Sec.~\ref{sec:evol}).
The left loop footpoint and the whole right leg are bright.  In this
filter the loop emission decreases with time as a whole.
The 284 \AA~filter profiles are very low. The left leg is never so
defined and bright as in the other two filter bands. The right leg
appears to be brighter in the first two profiles. Globally, the
profiles decay with time; the emission becomes negligible after 7:00
UT.
Also in the Yohkoh/SXT passband, the loop emission decreases with time.
The regions close to the footpoints appear to be both brighter than the
central loop region and the emission looks overall quite symmetric in
the first two profiles. Thereafter it may be compatible with zero
emission as in the TRACE 284 \AA~filter passband.

\subsubsection{Light curves}
\label{sec:lc}

\begin{figure}
\centerline{\psfig{figure=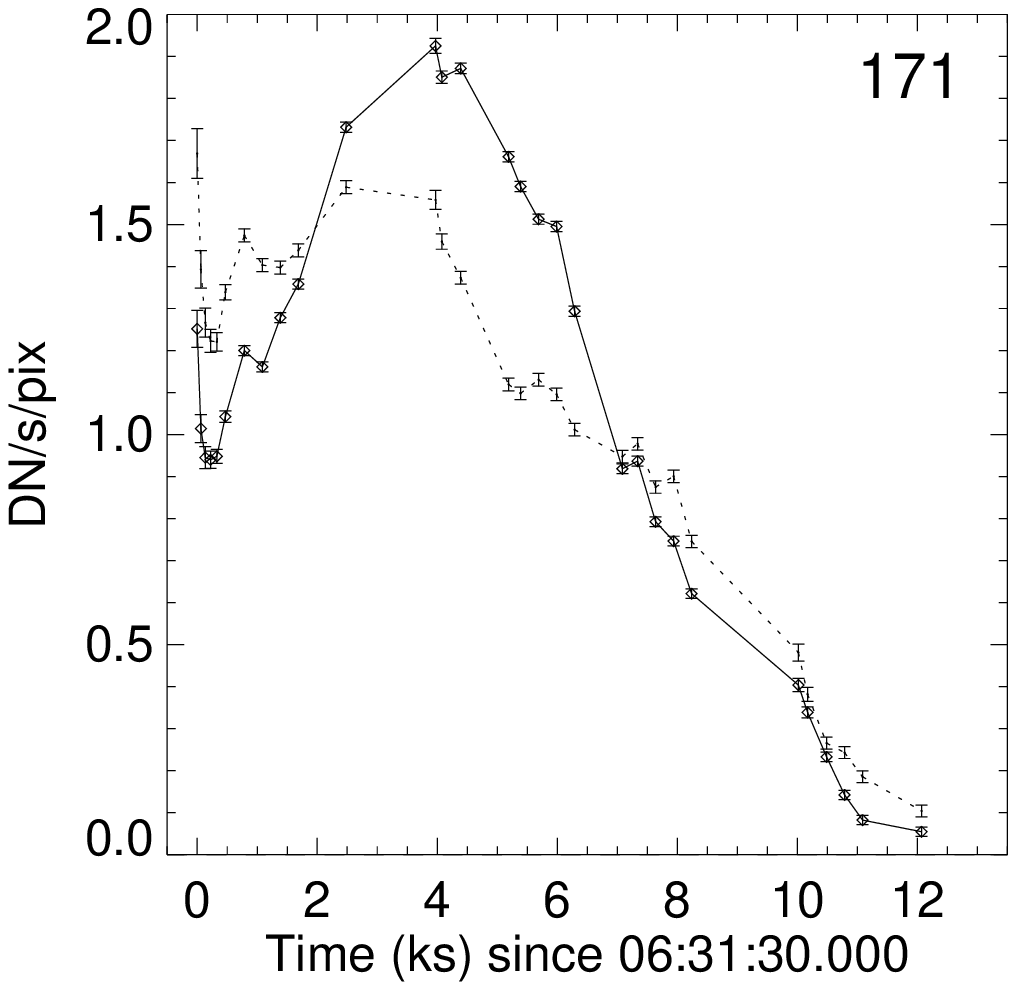,width=6cm}}
\centerline{\psfig{figure=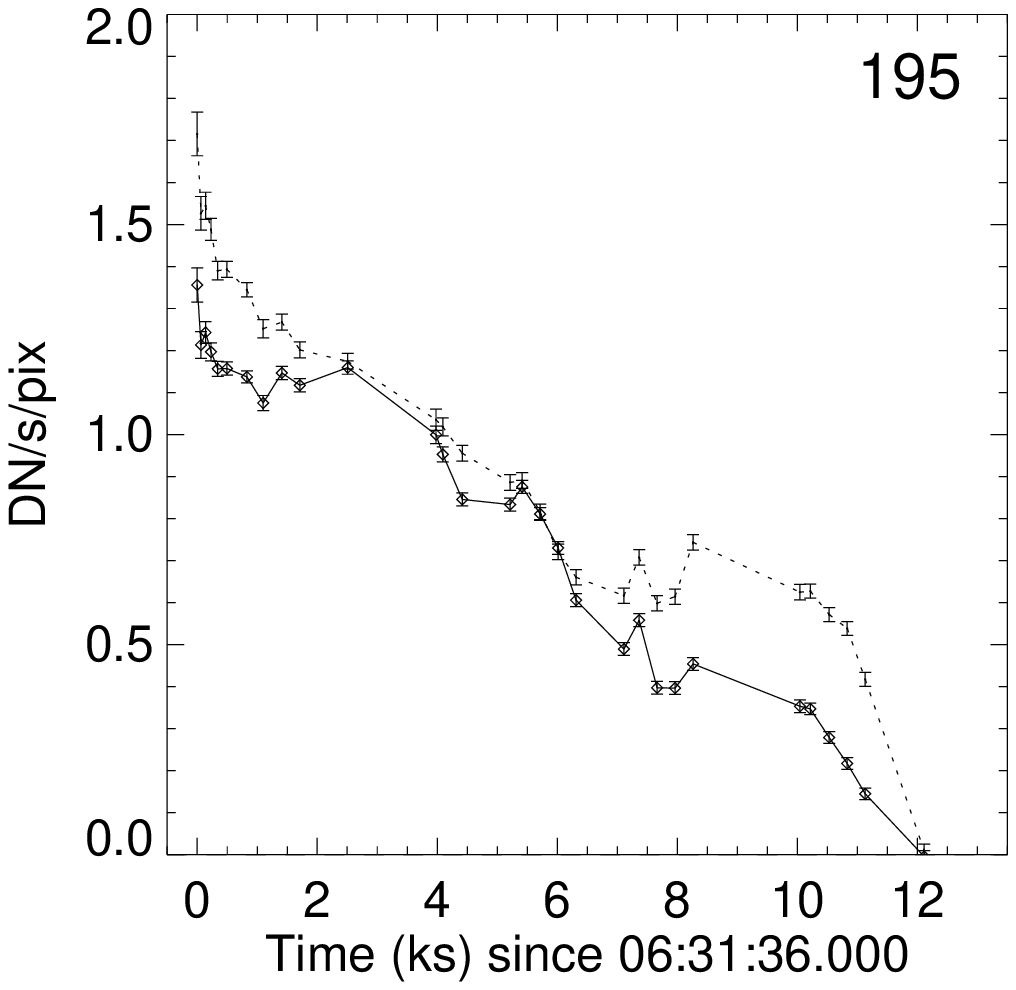,width=6cm}}
\centerline{\psfig{figure=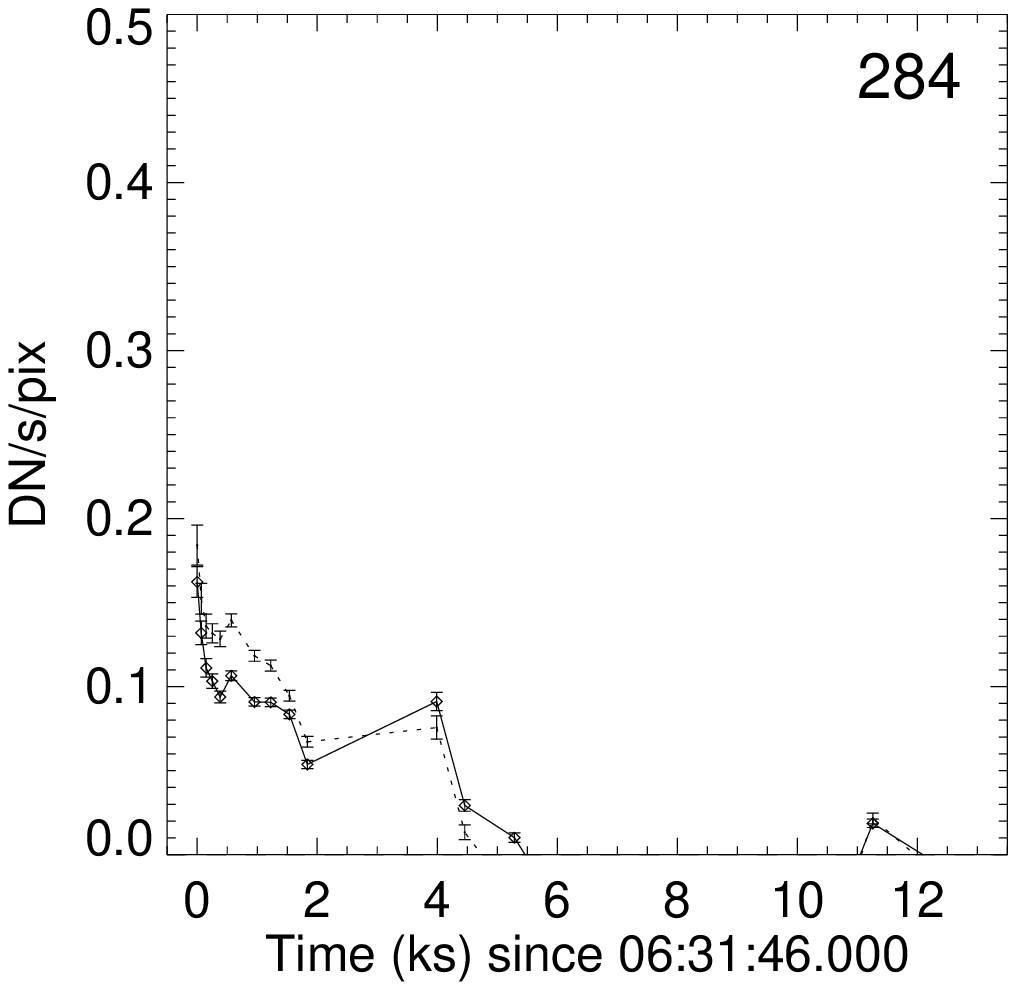,width=6cm}}
\centerline{\psfig{figure=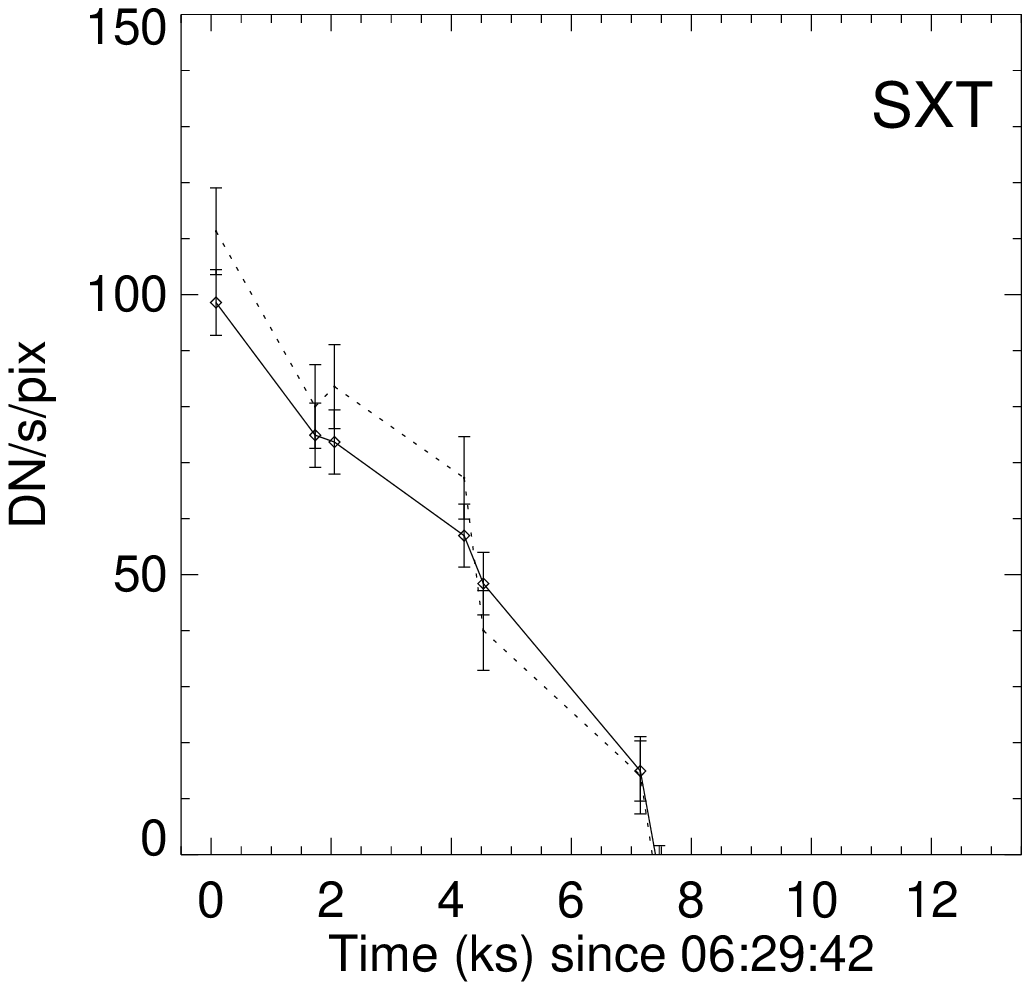,width=6cm}}
\caption[]{Background-subtracted 
light curves integrated along the whole loop strip (solid line)
and on the right loop half only (dashed line) 
in the TRACE 171 \AA, 195 \AA~and 284 \AA~
filter passbands, and in the Yohkoh/SXT Al/Mg/Mn filter passband.
\label{fig:lcbk}}
\end{figure}

We can build the light curve in the various filter passbands by
integrating the emission along the strips. The emission is integrated
from sector 4 to sector 26. For comparison we compute also the
evolution of the emission integrated only on the right half of the loop
(sector 14 to 26).
We will not show and comment on the unsubtracted light curves.  The
background-subtracted light curves both in the whole and in half of the
loop are shown in Fig.~\ref{fig:lcbk} (average DN rate per pixel).

In the TRACE 171 \AA~filter band, the light curve of the whole loop
first increases, with a peak about one hour after the beginning of
the observation, and then monotonically decreases to zero level at
$t \sim 12$ ks. The e-folding decay time of the whole loop is $\tau_{171}
= 2.5 \pm 0.4$ ks since the maximum.  Note that the free cooling time
(Serio et al. 1991) expected for a loop of half-length $L \approx 4.2
\times 10^9$ cm at 1 MK is about 1.6 ks, i.e. quite a shorter timescale.
The cooling times will be even shorter than this if the loops are either
over-dense or under-dense, as is typically the case for TRACE and
Yohkoh loops, respectively (e.g., Winebarger et al. 2003a)
The emission from the right leg appears more constant in the first part
of the observation, and then decreases as the total loop.

The light curve decays monotonically in all the other passbands, with
very similar trends in the whole and in the right half of the loop. 
The e-folding times are $\tau_{195} = 6.1 \pm 0.7$ ks,  $\tau_{284} =
2.9 \pm 0.7$ ks, $\tau_{SXT} = 4.0 \pm 0.7$ ks. The slowest decay is in
the 195 \AA~band, which is probably the one which detects most completely
the loop plasma cooling. In this band there are slight
increases of the emission of the whole and of the right half of the loop
at times $t \sim 2$ ks and $t \sim 8$ ks, respectively. In the TRACE
284 \AA~band and in the Yohkoh/SXT band, the decay is quite fast
and the emission becomes negligible after about 4 ks and 7 ks, respectively.

We have checked that the loop emission is relatively steady, with
fluctuations, in the four hours preceding the campaign, with an average
emission of 0.8 DN s$^{-1}$ pix$^{-1}$ in both the 171 \AA~and 195
\AA~filter bands, with a standard deviation of 0.2 and 0.1 DN s$^{-1}$
pix$^{-1}$, respectively. The beginning of the campaign approximately
corresponds to the time of maximum loop emission in the 195 \AA~band.
The loop emission in the SXT image at 5:26 UT is significantly lower
than the one at 6:29 UT (shown in Fig.~\ref{fig:fovbk}) approximately
by the same amount ($\sim 50$ \%) as the corresponding emission in the
TRACE 195 \AA~filter band.

\subsection{The TRACE filter ratios}
\label{sec:filt}

The temperature diagnostics from the ratio of the count rates in thin passband
filters such as those of TRACE suffer from possible uncertainties, because of
possible unpredictable variations of the element abundances of the few lines
emitted in the bands, of the temperature stratification along the line of
sight, and of the non-monotonic dependence of the filter ratio on the
temperature.

The problem is made even more difficult by the superposition of different
incoherent structures along the line sight. If this ``background"  is not
subtracted properly,  an artificial emission offset is created, which will
flatten the filter ratio profile along the structure, i.e. it will tend to make
the ratio closer to unity anywhere\footnote{Unless high contributions from cold
plasma are present.}. Therefore, one has to compute the filter ratio {\it
after} background subtraction. Although the loop data in the TRACE 284 \AA~
filter passband have some statistical significance in the initial part of the
observation, we have found the 284/195 filter ratio to yield large
uncertainties and we prefer not to comment on it. We have no way to obtain
background-subtracted images in two Yohkoh/SXT filter passbands (only in the
Al/Mg/Mn one) and we will address no filter ratio diagnostics for Yohkoh data.
We will therefore concentrate on the TRACE 195/171 filter ratio.

\begin{figure*}
\centerline{\psfig{figure=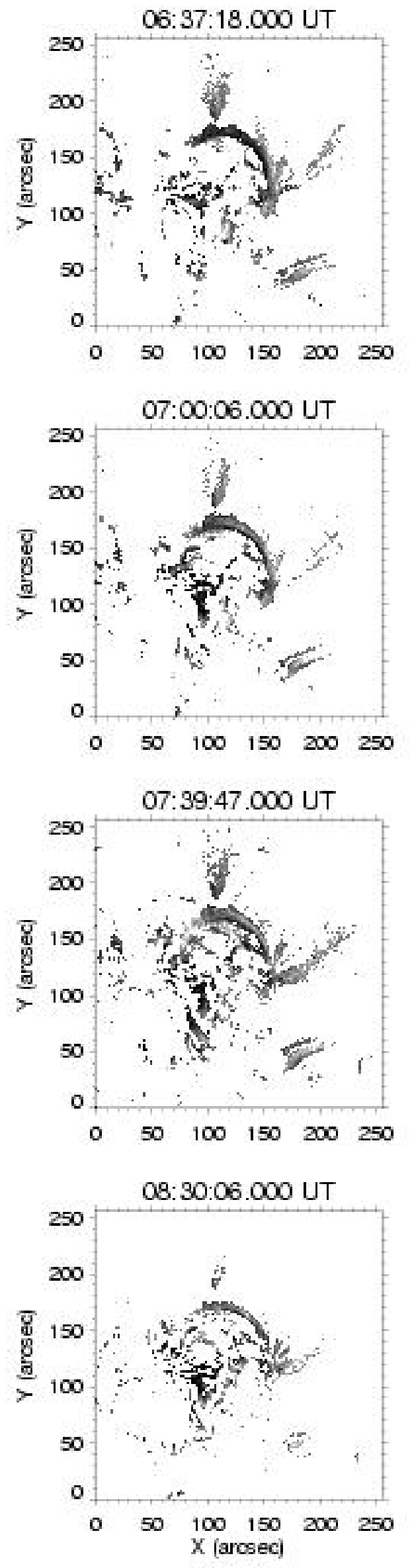,width=5.5cm}
\psfig{figure=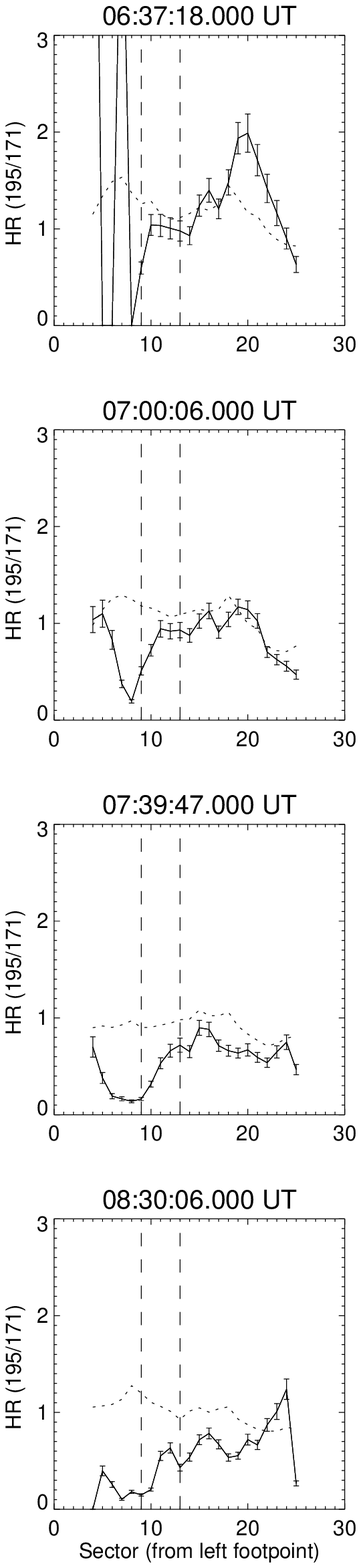,width=5.5cm}}
\caption[]{{\it Left}: Background-subtracted 195/171 filter ratio
maps of the loop region ($256 \times 256$ arcsec) from  6:30 UT to 8:30
UT.  The grey scale is logarithmic between 0.1 (white) and 2 (black).
The ratio is computed only where the pixel count rate is above 0.5 and
0.2 DN/s in the 171 \AA~and 195 \AA~filter bands, respectively.  {\it Right}:
Background-subtracted filter ratio 195/171 profiles along the loop.
Unsubtracted profiles are also shown ({\it dotted lines}).  }
\label{fig:fovhr}
\end{figure*}

As a first step, we obtain filter ratio maps, simply by computing the
pixel-to-pixel ratio of whole images, taken almost simultaneously (few seconds
difference) shown in  Fig.~\ref{fig:fovhr} at the same times as in
Fig.~\ref{fig:fovbk}.  The filter ratio is computed only where the pixel count
rate is above 0.5 and 0.2 DN/s in the 171 \AA~and 195 \AA~filter bands,
respectively, i.e. where there is enough S/N ratio.

Also thanks to the screening devised above, the loop shows up clearly in
the filter ratio maps, especially at early times, when the 195 \AA~signal
is high. This is an indication that the loop imaged by the
171 \AA~filter is well aligned to the loop imaged by the 195 \AA~filter,
supporting that both filters are detecting the same plasma.  The filter
ratio distribution appears full along the right leg, a gap (a zone of
low value) is present in the left leg.  In the chosen grey scale, darker
tones indicate higher filter ratio and, probably, hotter plasma. In the
initial frames, where the signal is higher, the filter ratio appears to
be structured, both across and along the loop. In the first map, higher
(darker) values are present in the inner shell and in the region of the
loop apex.  In the second and third map, both the outer and inner shells
of the loop appear to be darker. The loop color clearly shifts to paler
tones as time goes on, indicating that the structure is cooling, while
fading. The right footpoint appears to be at lower filter ratio values.

The right strip of Fig.~\ref{fig:fovhr} shows the filter ratio profiles at
the same times as the filter ratio maps (left strip).  All the profiles
are included along most of the loop in the range between 0.2 and 5, i.e.
$6 \le \log T \le 6.2$ (according to the standard TRACE software). The first
filter ratio profile has a peak around sector 20, i.e. in the middle of
the right half of the loop. This feature persists for about 5 min and
may be a local temperature maximum.
On the left loop side, there are low filter ratio values, forming
a well-defined dip. This dip is probably sharper than it should be,
because its right side is made steeper by the presence of the bright
intersecting structure. Given the expected filter ratio to temperature
relationship at temperatures below 1 MK (e.g. Lenz et al. 1999, Testa
et al. 2002), the dip may be the signature of a monotonic decrease of
the temperature from the loop top toward the left footpoint. However,
this indication is to be taken with care, since the filter ratio dip
is mostly caused by a dip in the 195 \AA~profile (Fig.~\ref{fig:profbk}),
which may be an artifact of the subtraction of a bright feature in that
region, as shown in Sec.~\ref{sec:evol}.
At intermediate times the filter ratio peak in the right loop side
disappears and the profiles are flatter there. At 8:30 UT, another
clear peak appears again on the right leg, closer to the footpoint.
Overall, the average emission level clearly progressively decreases.

For comparison, Fig.~\ref{fig:fovhr} also shows the profiles obtained
with no background subtraction. All unsubtracted profiles are flat and
equal, all around a value of filter ratio of one, and show no
evolution. 

\begin{figure}
\centerline{\psfig{figure=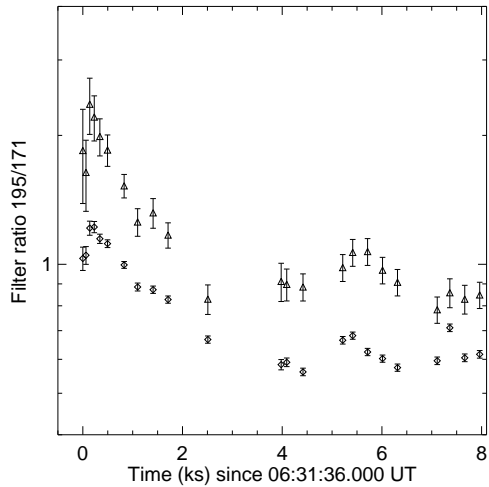,width=7.5cm}}
\caption[]{Time evolution of the filter ratio 195/171 averaged over the
loop (between sectors 8 and 25, diamonds) and the maximum ratio value taken
between sectors 13 and 20, i.e. the right leg of the loop (triangles).
\label{fig:hrlc}}
\end{figure}

Fig.~\ref{fig:hrlc} shows the evolution of the background-subtracted filter
ratio averaged over the loop (between sectors 8 and 25) and around the loop
apex (between sectors 13 and 20). The latter is systematically higher, due to
the non-uniform distribution of the filter ratio along the loop. Both ratios
clearly decrease with time: this is an evidence that the loop is gradually
cooling on average.

\subsection{The CDS data}

\begin{figure*}
\centerline{\psfig{figure=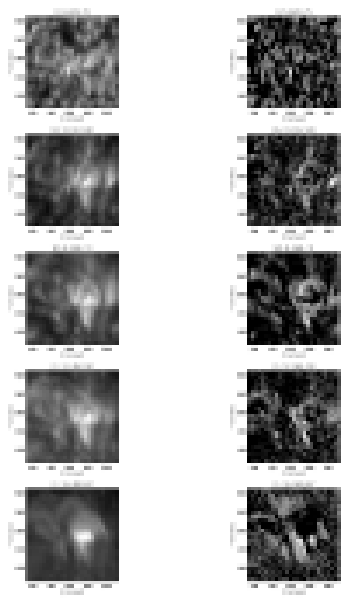,width=16cm}}
\caption[]{
Images of the loop region detected with SoHO/CDS in five spectral lines.
The left column show the images taken during the second raster from 
07:50:09 UT to 08:46:02 UT; on the right the same images are shown after the
background subtraction.
\label{fig:cds1}}
\end{figure*}

Figure~\ref{fig:cds1} shows the images in five spectral lines detected
with CDS during the first raster and the same images after the
background subtraction. The rastering roughly takes the time between the last
two panel rows in Fig.~\ref{fig:fovbk}.
In the difference images, the loop shows up in a few lines. It is best
visible in the Mg IX 368 \AA~line. Most of it is visible also in the Mg X
625 \AA~line and in the Si X lines (especially the 347 \AA~one). The left leg
is bright in the Ca X 558 \AA~line. The loop is barely visible in the Fe
XII 364 \AA~line, and not visible at all in the remaining lines. The hottest
lines of Fe XIV and Fe XVI may be detecting the right end. In summary,
we see the loop in the lines with peak formation temperature
in the range $5.9 \leq \log T \leq
6.1$, at best for $\log T = 6.0$, and we do not see it in the lines
outside this range. We also see the left leg as the brighter one. All
this is consistent with the loop as it appears in the 171 \AA~filter band
in the same time period (around 8 UT, see also
Fig.~\ref{fig:trace_cds}).

\begin{figure*}
\centerline{\psfig{figure=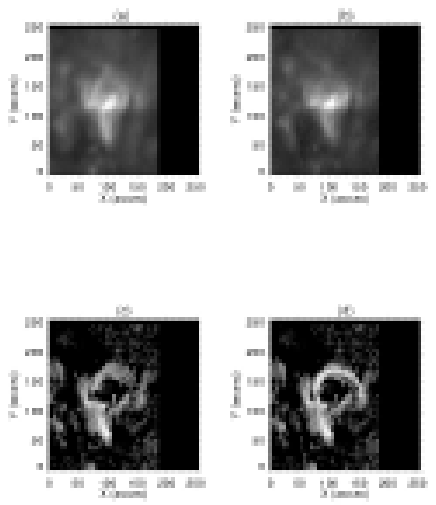,width=16cm}}
\caption[]{
The loop as seen by CDS in the Mg X 625 \AA~line. The the image of the loop is
in panel (a), the background image in panel is in panel (b).
Panel (c) shows the background subtracted image and panel (d) the same image
with the outline of the loop as obtained from TRACE images.
\label{fig:cds2}}
\end{figure*}

Figure~\ref{fig:cds2} shows a set of four images for the Mg X 625 \AA~line.
The upper row shows the images of the first and second raster (the background
image). The lower
row shows the difference images, the second one including the loop contour
as obtained from the TRACE data.  The coarser spatial resolution of CDS
images makes the loop less defined than in the TRACE images.

\begin{figure}
\centerline{\psfig{figure=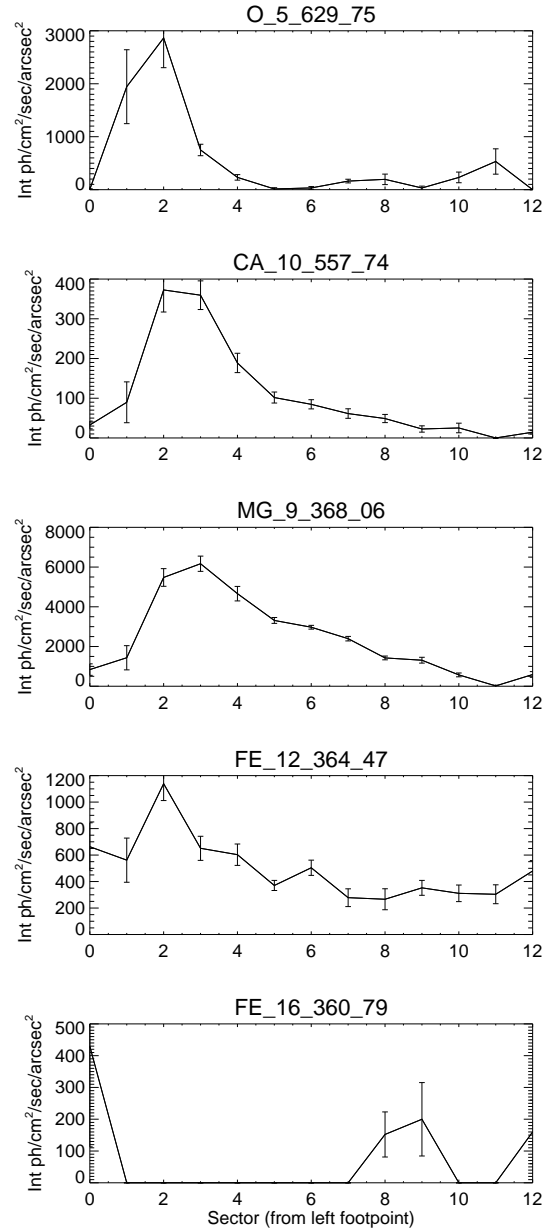,width=7.5cm}}
\caption[]{
Line intensities as function of the position (sector) along the loop. The 
loop has been divided in thirteen sectors as shown in Fig.~\ref{fig:cds2}.
\label{fig:cds3}}
\end{figure}

For the analysis of the emission from the loop plasma, in order to
match easily the TRACE data and to use the same loop outline, the CDS
images have been rebinned to the pixels size of TRACE. The loop has
been divided into 13 sectors in the rebinned difference images, and the
average line intensities have been computed in each sector. The loop
sectors in CDS images are different from -- approximately twice as long
-- those chosen to analyze the loop in the TRACE images
(Sec.~\ref{sec:loop}).  Few CDS sector pixels yield negative counts;
they have been put to zero.  Figure~\ref{fig:cds3} shows the intensities
in five lines along the loop. The error bars have been computed as
the standard deviation of the photon counts within each CDS sector.
The profiles show a well-defined trend in the lines where the loop
is visible in the difference images, i.e. from Ca X 558 \AA~to Fe
XII 364 \AA: there is an emission peak around sectors 1 to 3 and then
the emission decreases more or less gradually toward the right leg of
the loop. In the cool O V 629 \AA~line, the only significant feature is a
bump at the left extreme. There is no clear link of this bump to the loop
emission. The profile in the Mg IX 368~\AA~ line overall appears similar
to the background-subtracted profile in the TRACE 171 A filter band
at 7:39 UT (Fig.~\ref{fig:profbk}, see also Fig.~\ref{fig:trace_cds}).
In the hottest lines (Si XII, Fe XIV and Fe XVI) the profiles show weak
features on the right side of the loop region, which may be evidence of
some hot plasma there.  In the lines peaking around 1 MK, the profiles are
consistent with a more luminous left leg.  This appears to be consistent
with the emission distribution along the loop observed on average in
the 171 \AA~ filter band of TRACE in the corresponding time interval
(lower two plots in Fig.~\ref{fig:profbk}).

\subsubsection{Emission Measure}
\label{sec:em}

\begin{figure}
\centerline{\psfig{figure=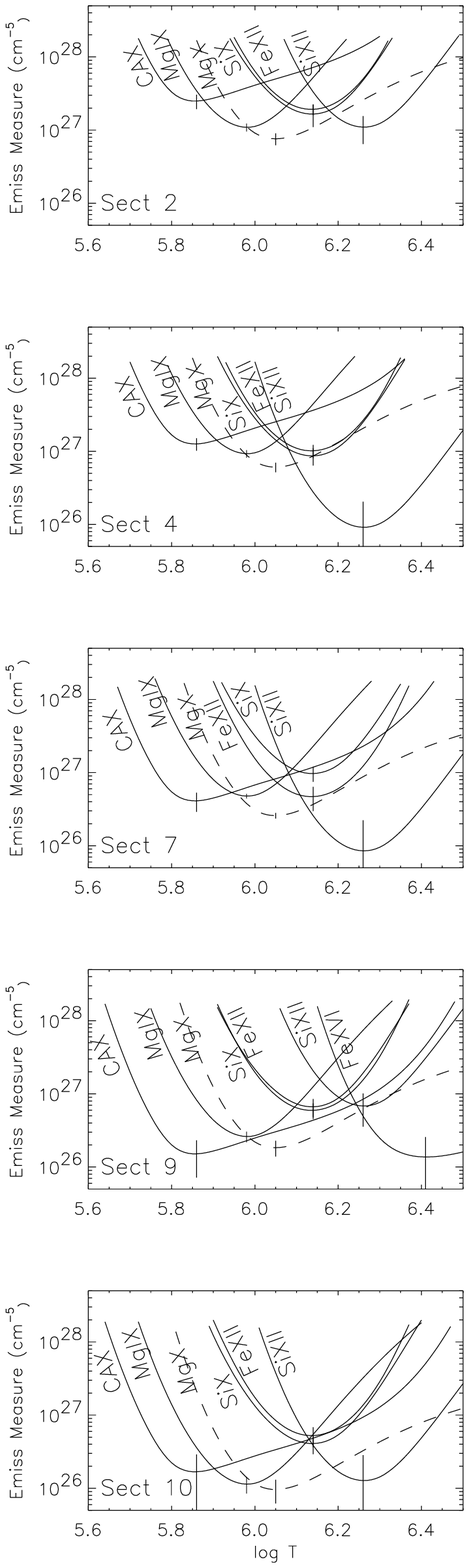,width=7.5cm}}
\caption[]{
Emission measure distribution in five CDS sectors (2, 4, 7, 9, 10)
of the loop.  The Mg X 625~\AA~ line ({\it dashed line}) appears to be
inconsistent with the others.}
\label{fig:cds4}
\end{figure}

Line emissivities were computed using CHIANTI V4.0.2 with ionization
equilibrium fractions from Mazzotta et al.  (1998) and photospheric
abundances.  Figure~\ref{fig:cds4} shows the emission measure loci
diagrams of five CDS sectors along the loop, namely CDS sectors 2, 4, 7, 9
and 10 (see Fig.~\ref{fig:cds3} for some of the line intensities),
roughly corresponding to TRACE sectors 4, 8, 14, 18 and 20
(Sec.~\ref{sec:loop}). CDS sectors 2 and 4 are in the left leg, CDS sector 7
around the loop apex, and CDS sectors 9 and 10 in the right leg.  The curves
are shown only for those lines in which the emission in the CDS sector
is larger than zero.
Among the lines we used in Fig.~\ref{fig:cds4}, Li-like (Mg X, Si XII)
and Na-like (Ca X) lines are those with the largest uncertainties.  As
shown by Landi et al. (2002a,b), the emission measures of Li-like lines
are lower by a factor of 2 when compared to the values obtained with other
spectral lines.  For the Na-like emissivities the status is less clear.
Na-like lines emissivities have been found to be higher by a factor of 2
(Landi et al 2000a) when compared with SUMER spectra. A good agreement
with the other ions has been found in comparison with CDS spectra
(Landi et al 2002b). We have found that the Mg X 625~\AA~line is
clearly inconsistent with all others in all sectors 
(dashed line in Fig.~\ref{fig:cds4}). 
Similar problems with the Mg X 625~\AA~line has been pointed out in
a specific spectral study made on solar data using the ADAS atomic and
spectral model (Lanzafame et al. 2005). We have then decided to discuss
the emission measure distributions without considering this line.

As a result, in Fig.~\ref{fig:cds4} CDS sector 9 appears to be the
hottest, with a peak around $\log T \approx 6.25$ and with a relatively
broad distribution on the cool side down to $\log T \approx 6.0$. CDS
sector 10 is somewhat cooler, well peaked around $\log T \approx 6.15$,
and CDS sectors 4 and 7 are even cooler with peaks at  $\log T \approx
6.05$, and again with a relatively broad distribution towaer the cool
side.  The situation of CDS sector 2 is less clear. This CDS sector is
located very close to the loop left footpoint and appears bright in the
Si XII line probably because of other bright overlapping structures. In
the light of this, the panel for CDS sector 2 may be compatible with
a relatively cool distribution at  $\log T \approx 6.1$, not different
from CDS sectors 4 and 7, and the Ca X line may suggest the presence of
significant cooler contributions.  This may be consistent with the bright
left leg of the loop in the TRACE 171 \AA~images (Fig.~\ref{fig:fovhr})
and in the cool lines CDS images (Fig.~\ref{fig:cds1}). Overall, the
figure suggests a trend of increasing temperature from the footpoints
to CDS sector 9, located in the right leg of the loop.

Knowing that the raster took about 1 h to span an X distance of 240",
we obtain that the time difference to raster from sector 4 to 10 (~50")
is about 12 minutes (0.7 ks). This is certainly too small a time lapse
to determine significant variations of this loop, and we conclude that
we are seeing the spatial structure of the loop. 

The emission measure appears to be the highest in CDS sector 2 ($\sim
2 \times 10^{27}$ cm$^{-5}$) and to decrease progressively toward the
right end of the loop ($\sim 5 \times 10^{26}$ cm$^{-5}$).

\section{Discussion}
\label{sec:disc}

\subsection{Data analysis and background subtraction}

In this work we present the analysis of a coronal loop imaged in multiple
filters and spectral lines over part of its lifetime.  The loop has been
selected on TRACE images. Our analysis collects the information coming
from the time-evolution of the loop region in three TRACE passbands,
and in one Yohkoh/SXT passband at overlapping times, and two SoHO/CDS
rasters in twelve relevant lines, one during the TRACE observation,
the other soon after. The selected loop is bright, lies on the disk and
part of it is well isolated from other bright structures. The fact that
we image a complete loop structure and monitor part of its evolution
provide further constraints of coherence for the data interpretation.

Since the loop fades out at the end of the TRACE and Yohkoh observations
and is absent in the second CDS raster, we could use the final TRACE
and Yohkoh frames and the second CDS raster as background to be
subtracted pixel-by-pixel from the other frames. The advantages and
disadvantages of this method of background subtraction are listed in
Section~\ref{sec:data}; a simple inspection of the background-subtracted
images supports that the method provides sound results.  The background
is indeed a significant fraction ($> 50$\%) of the total signal for all
instruments (Fig.~\ref{fig:profbk}), and its accurate subtraction is
therefore critical for any subsequent analysis and for determining any
physical parameter of a specific structure. In fact, we find differences
between indicators obtained from subtracted and unsubtracted data
(Fig~\ref{fig:hrlc}), the former being less uniform and more evolving.

\subsection{Loop morphology and evolution}

The loop is best visible and lives longer in the TRACE 171 \AA~and 195 \AA~
filter passbands. It is instead quite faint and decays rapidly in the
284 \AA~filter passband.

A simple superposition of the images, using the loop outlines, shows that the
loop in the 171 \AA~filter passband reasonably overlaps the loop in the 195
\AA~filter passband. The good correspondence is also supported by the loop
aspect in the 195/171 filter ratio maps. The alignment is not as good with the
loop in the 284 \AA~filter passband. The SXT loop is more diffuse: the right
leg appears to overlap that of the TRACE 171 \AA/195 \AA~loop. This may
indicate the presence of relatively hot plasma in the right leg of the loop.
The emission evolution in the SXT band appears to be quite coherent with that
in the TRACE 195 \AA~ band. However, because of the lower spatial resolution, no
conclusive statement can be made on the correspondence between the SXT and the
TRACE loop. With the same limitations, the TRACE loop overlaps well with the
loop as visible in a few SoHO/CDS spectral lines. The loop is best visible in
the Mg IX 368 \AA~line, but well visible also in the Mg X 625 \AA, in the Si X
347 \AA~and 356 \AA~lines, less in the Fe XII 364 \AA~ line.   All these lines
have temperature of maximum formation about $\log T \sim 6.0 - 6.1$, which is
also around the temperature of maximum sensitivity of both the TRACE 171
\AA~and 195 \AA~filter passbands.  Thus, there is a qualitative coherence of
the data from the different instruments. Limited parts of the loop seem to be
visible in a few spectral lines with other formation temperature, e.g. the left
leg in the cooler Ca X line ($\log T \approx 5.9$), the lower part of the right
leg in the hotter Fe lines.  This suggests that, during the first CDS raster,
in the right leg there is hotter plasma than in the left leg. This is
confirmed by the EM(T) reconstructed at five locations along the loop,
after removing the Mg X 625~\AA~ line.
In the cool O V line, we may be seeing only the footpoints of the loop.

We will not comment here about the structures surrounding the selected
loop.  We only mention that there is another bright structure which
apparently intersects the loop and which may have importance because it
appears to evolve coherently with the loop. This may be taken as
evidence of an interaction of this structure with the loop.

The TRACE images show that the loop is substructured in several strands,
as are many other loops observed with TRACE.  This work analyzes the loop
as a single and coherent structure, and we will not comment further on
the substructuring of the loop, which may deserve a further separate work.
We only note that TRACE appears to be detecting a thermal structuring
across the loop better than other instruments, as shown by the filter
ratio maps. Since anyway, the loop appears to have a coherent evolution,
the finer substructuring may not be crucial for the description of average
properties of the system.  The coherent evolution may suggest a coherent
heating across the structure.

TRACE and Yohkoh data allow us to analyze the evolution of the loop.
Unfortunately, CDS data are not able to provide analogous information
because there is only one relevant raster, i.e. a single snapshot,
during the loop evolution. High quality spectroheliograms intrinsically
require relatively long acquisition times.

\begin{figure*}
\centerline{
\psfig{figure=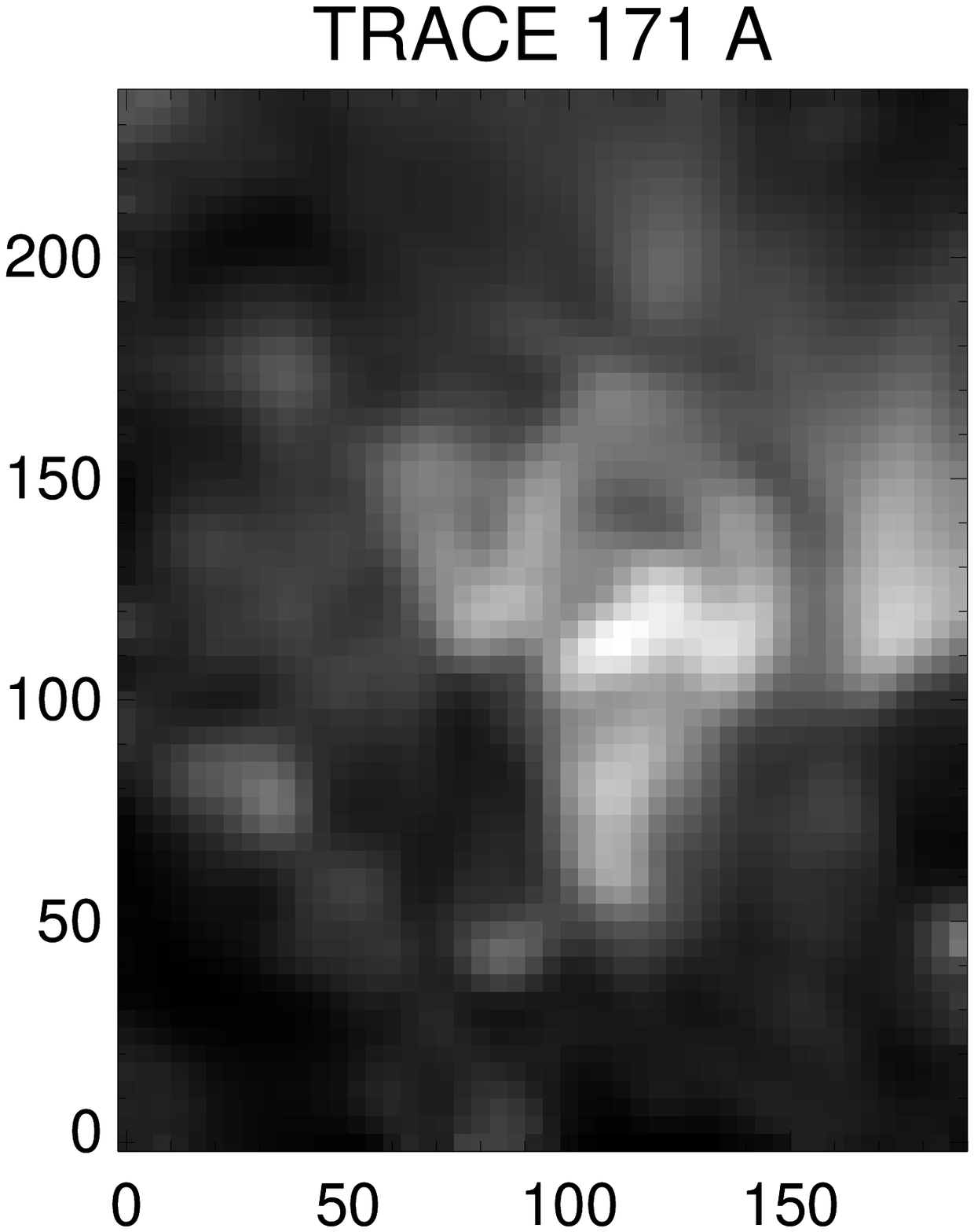,width=7cm}
\psfig{figure=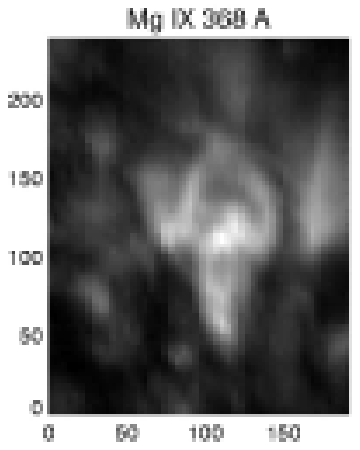,width=7cm}
}
\caption[]{
Image of the loop region obtained from the TRACE 171 \AA~images ({\it
left}) with a procedure which mimics the CDS rastering (at 
corresponding times) and degraded to
the CDS angular resolution (with a further smoothing with a $2\times2$
pixel boxcar), compared to the CDS (first) raster image in the Mg IX 368 \AA~line
({\it right}). The image coordinates are in arcsec.}
\label{fig:trace_cds}
\end{figure*}

In the analyzed time sequence the loop evolves. The TRACE 171 \AA~filter
detects an evolution both of the emission intensity and of its distribution
along the loop. In particular, the right leg of the loop is bright at the
beginning of the observation, while the left leg becomes the brighter later on.
The bright left leg is coherent with the brightness distribution in the
relevant CDS lines, taken in the same time period. Fig.~\ref{fig:trace_cds}
shows an image of the loop region obtained from the TRACE 171 \AA~images with a
procedure which mimics the CDS rastering (with corresponding time stepping) 
and degraded to the CDS angular resolution, compared to the CDS raster image in
the Mg IX 368 \AA~line.  The figure confirms a very good correspondence of the
loop appearance in both images. The asymmetry along the loop both in space and
in time may be an indication of an asymmetric distribution  of the heating
along the loop and a time variation of the heating intensity, to be checked
through detailed modeling. In the SXT and the other TRACE filters the emission
distributions is more uniform and evolves more uniformly. 

The SoHO campaign covers mostly the loop decay. This is clear from
the relevant light curves (shown in Fig.~\ref{fig:lcbk}): they have
a decreasing trend, with e-folding times between 0.7 and 1.7 h, with
the only exception of the TRACE 171 \AA~filter. In this filter, the
light curve has a peak about one hour after the observation starts and
then decays steadily.  This is coherent with other TRACE observations of
loops (Winebarger et al. 2003b). A coherent interpretation of the light
curves is compatible with a progressive cooling of the loop: the intensity
drops earlier in the filter passbands more sensitive to higher temperatures 
(SXT, TRACE 284 \AA). The peak in the 171 \AA~filter passband, which is
sensitive to cooler plasma than the others, may be simply due to the
progressive entrance of the cooling loop plasma into the temperature
regime of the highest filter sensitivity.  We have also checked
that the loop had been bright and relatively steady for quite a long time
(about four hours) before the start of the campaign.

The TRACE light curve decay times ($\sim 1$ hour) are significantly
longer that the characteristic cooling time expected for a loop of this
length from the equilibrium ($\sim 1/2$ hour or shorter). 
This may suggest that
the loop decay is sustained by a slowly decaying heating, as it often
occurs in flaring loops (e.g. Sylwester et al. 1993). We cannot exclude
other possible interpretations, e.g. an envelope of the free cooling
of different loop strands ignited at delay one from the other (Warren
et al. 2003) and/or significantly lower metal abundances. Detailed loop
modeling may shed further light on this topic.

\subsection{Loop physical conditions}

The pixel-by-pixel background subtraction allows us to derive TRACE 195/171
filter ratio maps. The maps show a moderate structuring of  the filter ratio
across the loop, with relatively higher filter ratio values in the inner  shell
and at the apex of the loop (Fig.~\ref{fig:fovhr}). The filter ratio maps also
show that the filter ratio has a coherent structuring along the loop and that
it evolves. This can be seen also in the background-subtracted filter ratio
profiles along the loop. The right leg of the loop does not intersect with
other bright structures along the line of sight and the filter ratio
diagnostics seem to be more reliable there. We see a filter ratio peak there in
the initial profiles, which may be the signature of a temperature maximum.  In
the left leg the profiles show a deep minimum, which may be taken as a
signature of monotonic decrease of the temperature toward the left footpoint.
The evidence is not conclusive, because of the presence of a bright co-evolving
structure which intersects along the line of sight. The filter ratio profiles
along the loop are moderately non-uniform. Other transient ratio peaks
sometimes occur mostly at the right visible footpoint. Although with caution,
we suggest that these peaks may be the signature of minor heating episodes.

The 195/171 ratio values are mostly contained in the range 0.1-5, which should
correspond to a temperature range $6.0 \le \log T \le 6.2$.  Initially, the
higher ratio values and the good visibility in the TRACE 195~\AA~ band indicate
the presence of relatively hotter plasma than later, with better evidence in
the right leg of the loop. The right leg of the loop is also relatively bright
at early times in the 284~\AA~band and in the SXT band, suggesting the presence
of even hotter plasma in this phase and, therefore, a moderately multi-thermal
structure across the loop. The absence of spectral data in this phase does not
allow us to confirm this indication. The SXT and TRACE 284~\AA~light curves
show however that such hotter plasma  rapidly cools down and coherently is no
longer present during the CDS raster. We may conjecture the presence of hot
strands at the beginning of the campaign which either decay more rapidly than
the others or become uniform to them; this may mean that in the life of the
same loop we may find more and less  multi-thermal phases at different times.
This may not be in disagreement with the presence of both multi-thermal and
isothermal cross-structures suggested by Schmelz et al. (2005) and represent a
further piece in the puzzle. Detailed modeling, and, more probably, next
generation multi-band imaging observations will help to clarify this important
issue.

Later, during the first CDS raster, the loop probably overall cools closer to
$\log T \approx 6.0-6.1$.  This temperature regime is largely coherent with the
loop appearance in the spectral lines of SoHO/CDS: the loop is best visible in
the Mg IX 368 \AA~line, which peaks just at $\log T \approx 6.0$, and
we see the loop also in the Mg X 625 \AA, Fe XII 364 \AA, and in the
Si X lines ($\log T \approx 6.1 $). The loop is instead hardly visible
in the lines just outside this temperature regime. From combining CDS
data with theoretical line emissivities, we can sample emission measure
distributions along the loop. After removing the Mg X 625~\AA~ line, 
clearly inconsistent with all others, we obtain that the hottest components lie
on the right side of the loop (CDS sector 9), with a peak temperature of
$\log T \approx 6.25$ (Fig~\ref{fig:cds4}). The emission measure drops
sharply above this temperature, and generally more smoothly to the cool
side. At this time, on the right leg, the TRACE 195/171 filter ratio
is around 1, corresponding to a temperature $\log T \approx 6.1$. This
may suggest that, in this moderately multi-thermal structure, TRACE
is more sensitive to the plasma at intermediate temperature. It will
be interesting to check this hypothesis through detailed loop forward
modeling.  Assuming a loop thickness $\sim 5 \times 10^8$ cm an emission
measure of $\sim 10^{27}$ cm$^{-5}$ is compatible with a density $\sim
1.5 \times 10^9$ cm$^{-3}$, moderately overdense with respect to a loop at
temperature $\log T \approx 6.2$ (according to the scaling laws of Rosner
et al. 1978), consistently with the fact that the loop is cooling. 

The TRACE filter ratio values averaged along the loop (Fig.~\ref{fig:hrlc})
definitely decrease with time, supporting the progressive cooling of the loop.
The decrease somewhat slows down with time, perhaps indicating the occurrence
of secondary heating episodes, which would by the way be consistent with the
observed long decay time of the loop.

\section{Conclusions}
\label{sec:concl}

In this work we use a good combination of multi-wavelength data,
time/space/spectral resolution and signal-to-noise ratio to investigate the how
deep one can go in the direct diagnostics and interpretation of a coronal
observation. The observation of a specific structure, a coronal loop, and of
its evolution helps us in the attempt to find coherent results and 
limitations to the information that it is possible to derive.

The analysis of the collected information shows that it is overall possible to
obtain a coherent scenario and to  detect several details of the emission and
evolution of a coronal loop. It confirms temperature diagnostics with TRACE to
be difficult, and a proper subtraction of the high background critical for it, 
but also indicates that, in particular conditions, some sound information can
be obtained. Spectroscopic data from SoHO/CDS provide useful complementary
information, constraining the loop thermal structure, although in the limit of
lower temporal and spatial resolution.

The coherent scenario that we obtain across bands and instruments appears
to confirm the overall evolution of the coronal loops entirely visible
with TRACE (Warren et al. 2002, 2003) and the presence of thermal structuring
(Schmelz et al. 2005), but also adds several qualitative
and quantitative details and puts several constraints to be matched
coherently through detailed loop modeling.

\acknowledgements{We thank J. C. Raymond, H. Hudson, J. Klimchuk and
A. Maggio for suggestions.  
The authors
acknowledge support for this work from Agenzia Spaziale Italiana and
Ministero dell'Istruzione, Universit\`a e Ricerca.}

\end{document}